\documentclass[journal,11pt,onecolumn,a4paper]{IEEEtran}
\usepackage{graphicx}
\usepackage{amssymb}
\usepackage{amsmath}
\usepackage{cite}
\begin{document}

\title{AC losses  in type-II superconductors induced by nonuniform
fluctuations of external magnetic field}
\author{Leonid~Prigozhin and Vladimir~
Sokolovsky%
\thanks{L. Prigozhin is with the Department of Solar Energy and
Environmental Physics, Blaustein Institute for Desert Research,
Ben Gurion University of the Negev, Sede Boqer Campus, 48990
Israel (e-mail: leonid@cs.bgu.ac.il)}%
\thanks{V. Sokolovsky is with the Physics Department, Ben Gurion University of the
Negev, Beer Sheva, 84105 Israel (e-mail:
sokolovv@bgumail.bgu.ac.il)}} \maketitle
\begin{abstract}
Magnetic field fluctuations are inevitable in practical
applications of superconductors and it is often necessary to
estimate the AC losses these fluctuations induce. If the
fluctuation wavelength is greater than the size of a
superconductor, known estimates for an alternating uniform
external magnetic field can be employed. Here we consider the
opposite case and analyze, using a model critical-state problem,
penetration of spatially nonuniform fluctuations into type-II
superconductors. Numerical simulation is based on a variational
formulation of the Bean model. The analytical solutions, found in
a weak penetration limit, are used to evaluate AC losses for two
types of fluctuations: the running and standing waves. It is shown
that for spatially nonuniform fluctuations the losses are better
characterized by the fluctuation penetration depth than by the
fluctuation amplitude. The results can be used to estimate the AC
losses in flywheels, electric motors, magnetic shields, etc.
\end{abstract}
\begin{keywords}
Hard superconductors, Bean model, AC losses, penetration depth,
nonuniform field fluctuations, asymptotic solution.
\end{keywords}
\section{Introduction}
 The possible range of currents and magnetic fields as
well as the economic gains of implementation of type-II
superconductors in power transmission lines, current leads, fault
current limiters, magnetic shields, bearings, etc. are often
limited by AC losses and the necessity to remove the generated
heat out of the system. Thus, application of bulk high-Tc
superconductors in flywheel systems and  magnetic bearings is
promising because of no friction between the moving parts and,
hence, no energy losses by friction. However, rotating permanent
magnets used in such devices always produce somewhat irregular
magnetic field; moving field irregularities cause  hysteretic
losses and relaxation of levitation property \cite{Takase}.

The mathematical models, used to analyze magnetization of type-II
superconductors and to evaluate AC losses, involve highly
nonlinear partial differential equations that have been solved
mostly for superconductors in a uniform alternating external
magnetic field (see \cite{Brandt1,Brandt2,IEEE,Maslouh,SM} and the
references therein). In many practical situations, however, the
external magnetic field is not exactly uniform and can be better
presented as a superposition of a uniform part and spatiotemporal
fluctuations with the characteristic length scale less than the
size of a superconductor. The fluctuations are often stochastic
but, for example, in the case of a flywheel system, are induced by
rotation of permanent magnets and may be approximated by a running
wave. If there are no moving parts as, e.g., in the case of
magnetic shields or transformers, nonuniform magnetic field
fluctuations in the form of a standing wave can, probably, serve
as a better approximation.

Our aim is to study the penetration of nonuniform magnetic field
fluctuations into a hard superconductor and to evaluate the
accompanying AC losses. We start with a convenient for numerical
simulations variational general formulation of the Bean
critical-state model (section \ref{var_form}), then consider the
simplest geometric configuration, a superconductive slab placed
between two parallel sheets of external current. Even in this
simplest case the problem becomes nontrivial if the external
current and, hence, also the external magnetic field, are
nonuniform. We assume the external field fluctuations are induced
by a given current in the form of either a running or a standing
wave and solve, first, the magnetization problems numerically
(section \ref{2types}). Simple physical arguments allow us to find
also asymptotic analytical solutions for small fluctuations
(section \ref{Asimp_sec}). We further extend these solutions by
presenting them  as the zero-order terms of consistent asymptotic
expansions, find the first order corrections (appendices I and
II), and, finally, determine the asymptotic AC losses for small
fluctuations (section \ref{ACloss}). We analyze the dependance of
the leading AC loss term on the first order correction to current
density distribution.

\section{Variational formulation of the Bean model\label{var_form}}
Let a superconductor occupying domain $\Omega\subset \mathbb{R}^3$
be placed into magnetic field $\mathbf{H}_e(r,t)$ induced by a
given external current with the density $\mathbf{J}_e(r,t)$ (here
$r=\{x,y,z\}$ and $\mathbf{J}_e=0$ in $\Omega$). In accordance
with the Faraday law, an alternating magnetic flux induces
electric field and, hence, an eddy current inside the
superconductor. In an ordinary conductor, the vectors of the
electric field and current density are related by the linear Ohm
law. Type-II superconductors are, instead, characterized in the
Bean critical state model \cite{Bean} by a highly nonlinear
current-voltage relation which gives rise to a free boundary
problem.

Let us assume
\begin{equation}\mathbf{E}=\rho \mathbf{J}\label{Ohm}\end{equation}
and employ the usual Bean model relations determining the
effective resistivity of a superconductor, $\rho(r,t)$,
implicitly. Namely, the Bean model states that the effective
resistivity is nonnegative,
\begin{equation}\rho(r,t)\geq 0,\label{rho}\end{equation} the
current density cannot exceed some critical value,
\begin{equation} |\mathbf{J}(r,t)|\leq
J_c,\label{Jc}\end{equation} and, if the current density is
subcritical, the resistivity is zero:
\begin{equation} |\mathbf{J}(r,t)|<
J_c\rightarrow\rho(r,t)=0.\label{cond}\end{equation}

\noindent Since no current is supposed to be fed into the
superconductor by an electric contact, the current density inside
$\Omega$ should satisfy the zero divergence condition and have a
zero normal component at the domain boundary $\partial\Omega$:
\begin{equation} \nabla\cdot\mathbf{J}=0\ \mbox{in}\ \Omega,\ \
\ \ \ \ \mathbf{J}_n=0\  \mbox{on}\
\partial\Omega.\label{div0}\end{equation}
 To derive a
variational formulation of the magnetization problem, we define
the set $K$ of possible current densities,
$$\mathbf{J}\in K=\left\{\mathbf{\Phi}(r,t)\ \left|\begin{array}{lr}
\nabla\cdot\mathbf{\Phi}=0\ & \mbox{in}\ \Omega,\\
\mathbf{\Phi}_n=0\ & \mbox{on}\ \partial\Omega,\\
|\mathbf{\Phi}|\leq J_c\ & \mbox{in}\ \Omega\
\end{array}\right.\right\},
$$
and express the electric field via the vector and scalar magnetic
potentials,
\begin{equation}\mathbf{E}=-\partial_t\mathbf{A}-\nabla \psi
.\label{E}\end{equation} We further eliminate the scalar potential
by multiplying  (\ref{E}) by $\mathbf{\Phi}-\mathbf{J}$,
integrating over $\Omega$, and making use of the zero divergence
condition for functions from the set $K$, \begin{equation}
\left(\mathbf{E},\mathbf{\Phi}-\mathbf{J}\right)=-\left(\partial_t\mathbf{A},
\mathbf{\Phi}-\mathbf{J}\right),\label{EA}\end{equation} for
$\mathbf{J},\mathbf{\Phi}\in K.$ Here
$(\mathbf{u},\mathbf{v})=\int_{\Omega}\mathbf{u}\cdot\mathbf{v}\,d\Omega$
is the scalar product of two functions.

Since $\mathbf{E}$ and $\mathbf{J}$ satisfy the current-voltage
relations of the Bean model, the scalar product on the left side
of  (\ref{EA}) is nonpositive for any $\mathbf{\Phi}\in K$.
Indeed, using (\ref{Ohm})-(\ref{cond}) we obtain: \begin{equation}
\mathbf{E}\cdot\mathbf{J}=|\mathbf{E}||\mathbf{J}|=|\mathbf{E}|J_c\geq
|\mathbf{E}||\mathbf{\Phi}|\geq\mathbf{E}\cdot\mathbf{\Phi}.\label{EJ}
\end{equation}
Up to the gradient of a scalar function, determined by the gauge
and also eliminated by the scalar product with
$\mathbf{\Phi}-\mathbf{J}$, the vector potential is a convolution
of the Green function of Laplace equation, $G(r)=1/4\pi|r|$, and
the total current density $\mathbf{J}(r,t)+\mathbf{J}_e(r,t)$:
$$\mathbf{A}=\mu_0 G*(\mathbf{J}+\mathbf{J}_e)$$
(it is assumed that the magnetic permeability of superconductor is equal to
that of vacuum, $\mu_0$.) We arrived at the following variational
formulation of the magnetization problem: \begin{equation}
\begin{array}{c} \mbox{Find}\ \mathbf{J}\in K\ \mbox{such that}\\
\left(G*\partial_t\{\mathbf{J}+\mathbf{J}_e\},\mathbf{\Phi}-\mathbf{J}\right)\geq
0\ \ \mbox{for any}\ \mathbf{\Phi}\in K,\\
\mathbf{J}|_{t=0}=\mathbf{J}_0(r),\end{array} \label{vi}\end{equation}
where
$\mathbf{J}_0\in K$ is a given initial distribution of the current
density.

The problem (\ref{vi}) is an evolutionary variational inequality
with a nonlocal operator and has a unique solution \cite{EJAM}.
This inequality is written for the induced current density alone:
the effective resistivity $\rho$ has been used to derive the
inequality (\ref{EJ}) and then excluded. It has been shown
\cite{EJAM} that, in the Bean model, $\rho(r,t)$ is the Lagrange
multiplier related to the current density constraint (\ref{Jc}).
Of course, the same inequality may be written as
$\left(G*\partial_t\mathbf{J}+\frac{1}{\mu_0}\partial_t
\mathbf{A}_e,\mathbf{\Phi}-\mathbf{J}\right)\geq
0$ for any $\mathbf{\Phi}\in K$, where $\mathbf{A}_e$ is the
external vector potential; such formulation can be more convenient
in some cases.

We note that, although the variational formulations where the
solution is sought as an extremal point of some functional are
much more familiar, the variational inequalities do appear in many
problems of mechanics and physics containing a unilateral
constraint or a non-smooth constitutive relation (see \cite{DL}).
The methods for numerical solution of variational inequalities are
well developed \cite{GLT}. Nevertheless, solution of (\ref{vi}) in
the general 3d case is certainly a challenging problem not only
because of huge number of finite elements needed: the nonlocal
operator of convolution leads to a dense matrix of the discretized
problem, and the zero divergence condition (\ref{div0}) is not so
easy to account for numerically, although the edge finite elements
and tree-cotree decomposition can be helpful (see
\cite{ARB,BossavitB}). Similar difficulties are typical also of
 other formulations of the magnetization problem
\cite{Bossavit,Brandt1}. That is why the magnetization problems
were solved mainly for one and two-dimensional configurations.
Below, we use a model 2d problem to simulate the penetration of
nonuniform magnetic field fluctuations into a hard superconductor
numerically and to calculate the AC losses asymptotically for weak
penetration.

\section{Two types of field fluctuations\label{2types}}
To simulate the penetration of magnetic field fluctuations into a
superconductor, let us consider  the simplest possible
configuration: a superconductive slab $-l\le x\le l$ placed
between two sheets of external current at $x=\pm a$, $a>l$ (Fig.
1).We assume the external current is parallel to $z$-axis,
$\mathbf{J}_e=J_e{\mathbf e}_z$.
\begin{figure}[h]\label{FigSlab}\begin{center}
\includegraphics[width=4cm,height=5cm]{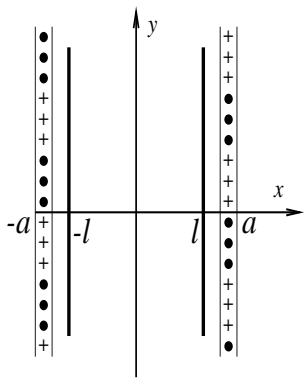}
\caption{Infinite slab between two parallel sheets of
current.}\end{center}\end{figure}  Having in mind, e.g., a
superconductor moving along the axis of a solenoid with slightly
nonuniform winding, we present the external current $J_e$ as a sum
of its uniform component $I_e$ and a wave moving along the
$y$-axis with the velocity $v$:
\begin{equation}J_{e}=\left[I_e+i_{e} \sin
\{
k(y-vt)\}\right][\delta(x+a)-\delta(x-a)].\label{JeV}\end{equation}
We will also consider the standing wave of  magnetic field induced
by the current
\begin{equation}J_{e}=\left[I_e+i_{e} \sin
(ky)\sin(2\pi f
t)\right][\delta(x+a)-\delta(x-a)].\label{JeS}\end{equation}

If the sheet current density were uniform, the Bean model
equations for slab magnetization could be solved easily. It is
more difficult to analyze the magnetization in a nonuniform
magnetic field induced by the external currents (\ref{JeV}) or
(\ref{JeS}), and we will use a 2d reformulation of the variational
inequality (\ref{vi}) and the numerical method proposed in
\cite{IEEE} for cylinders of arbitrary cross sections in a
perpendicular magnetic field. The method can be employed for any
distribution of  external currents. Taking the cylinder cross
section to be a rectangle, $-l\le x \le l,\ L\le y\le L$, we model
the development near the slab surfaces of critical current zones
that shield the fluctuating component of external magnetic field.

The shielding  current inside the superconductor is directed along
the $z$-axis and does not depend on $z$,
$\mathbf{J}=J(x,y,t)\mathbf{e}_z$, so the conditions (\ref{div0})
are satisfied automatically. We may redefine $\Omega$ to be the
cylinder cross section and solve a 2d problem in this domain. One
should, however, be cautious: the eddy current running in the
positive $z$-axis direction has to return back (no transport
current is applied; see \cite{IEEE} for solution of problems with
transport current). As a trace of the 3d zero-divergence
conditions (\ref{div0}), the condition
$$\int_{\Omega}J(x,y,t)\,d\Omega=0$$ has to be satisfied for all $t$.
The set of admissible (scalar) current densities becomes
$$K=\left\{\Phi(x,y,t)\ \left|\ |\Phi|\leq J_c \ \mbox{in}\ \Omega,\ \
\int_{\Omega}\Phi\,d\Omega=0\right.\right\}.$$ The variational
inequality (\ref{vi}) can also be rewritten for scalar current
densities,
\begin{equation}
\begin{array}{c} \mbox{Find}\ {J}\in K\ \mbox{such that}\\
\left(G*\partial_t\{{J}+{J}_e\},{\Phi}-{J}\right)\geq
0\ \ \mbox{for any}\ {\Phi}\in K,\\
{J}|_{t=0}={J}_0(r),\end{array} \label{vi1}\end{equation} where
$G(r)=-\frac{1}{2\pi}\ln |r|$ is now the Green function of Laplace
equation in $\mathbb{R}^2$
 and $r=\{x,y\}$.

To solve the problem numerically, we use first the finite
difference discretization in time and obtain, for each time layer,
the stationary variational inequalities
$$\mbox{Find}\ {J}\in K\ \mbox{such that}\ \
(G*\{J+J_e-\widehat{J}-\widehat{J}_e\},\Phi-J)\ge 0\ \ \mbox{for
any}\ \Phi\in K,$$ where $\widehat{J},\widehat{J}_e$ are the
values from the previous time layer. It can be shown that these
variational inequalities are equivalent to the optimization
problems \begin{equation}\begin{array}{cc}\min &
\frac{1}{2}(G*J,J)+(g,J),\\J\in K&\
\end{array}\label{opt}\end{equation}
where $g=G*\{J_e-\widehat{J}-\widehat{J}_e\}$. The convolution
$G*J_e$ can be calculated analytically:
\begin{eqnarray} \frac{1}{\mu_0}A_e=G*J_{e}=-I_ex
-\frac{i_{e}}{k}e^{-ka} \sin\{ k(y-vt)\}\sinh(kx)\label{AeV}\\
\frac{1}{\mu_0}A_e=G*J_{e}=-I_ex -\frac{i_{e}}{k}e^{-ka}\sin(2\pi
f t)\sin(ky)\sinh(kx) \label{AeS}\end{eqnarray} for currents
(\ref{JeV}) and (\ref{JeS}), respectively, and $-a<x<a$.

We finally discretize the optimization problems (\ref{opt}) in
space by means of the piecewise constant finite elements, take
care of the integral constraint $\int_{\Omega}J\,d\Omega=0$ using
the Lagrange multipliers technique, and solve the resulting
finite-dimensional optimization problems with  the remaining
constraints by a point relaxation method (see \cite{IEEE,JCP} for
the implementation details).

Let us assume only an alternating part of the external current is
present and $I_e=0$. A permanent field would not change the
picture qualitatively and, although in applications the stationary
component is often much stronger than fluctuations\footnote{Note
that strong permanent component makes  the relative changes of
magnetic field small. This justifies the use of field-independent
critical current density for simulating the penetration of
fluctuations. Otherwise, the Kim model \cite{Kim} or its
modification can be employed.}, the constant part causes no energy
losses. We present the results of two simulations. In both cases
we assume the superconductor is initially in the virgin state,
$J_0=0$.

\begin{figure}[t!]
\centering
\includegraphics[width=3.cm,height=8cm]{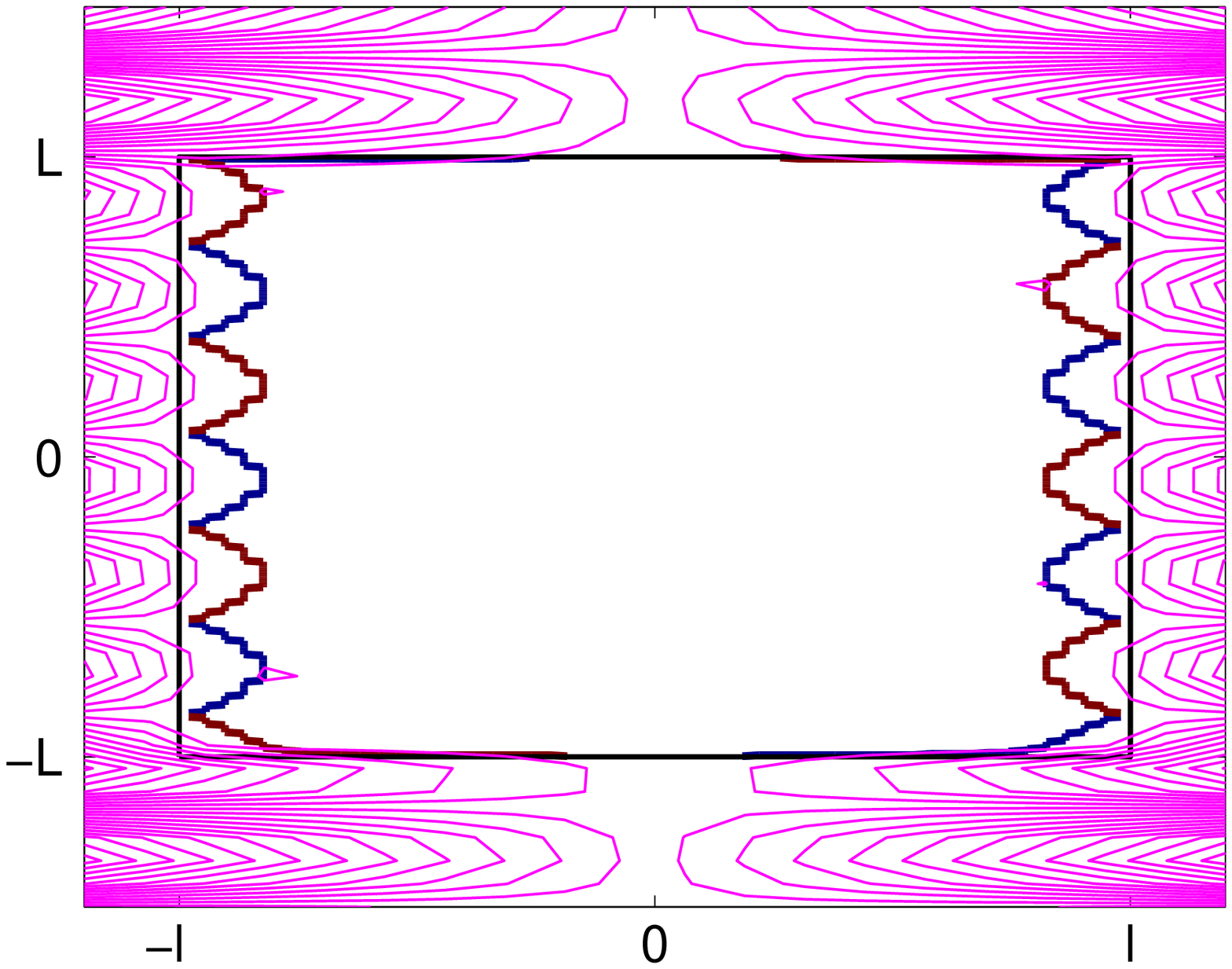}\hspace{.1cm}
\includegraphics[width=3.cm,height=8cm]{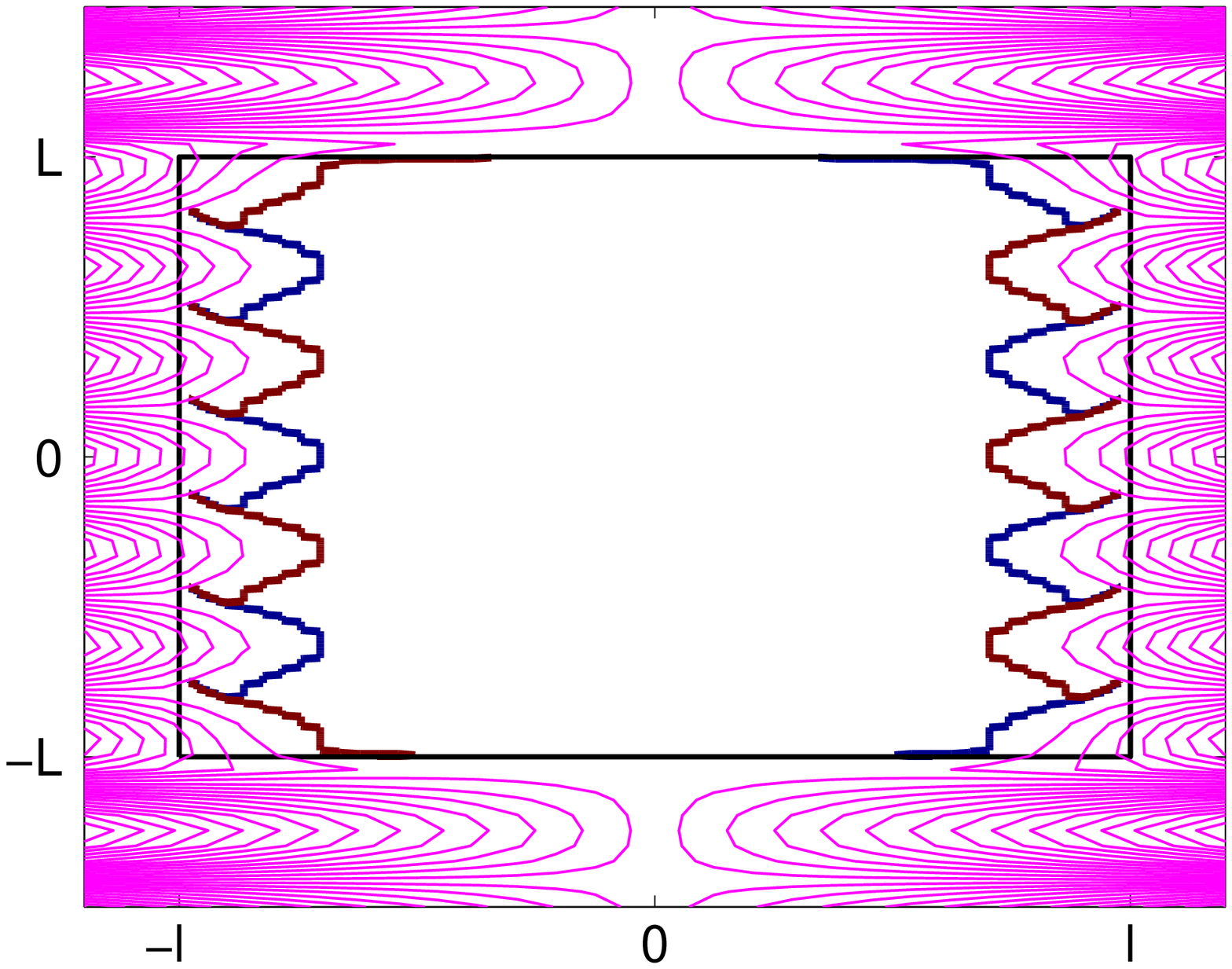}\hspace{.1cm}
\includegraphics[width=3.cm,height=8cm]{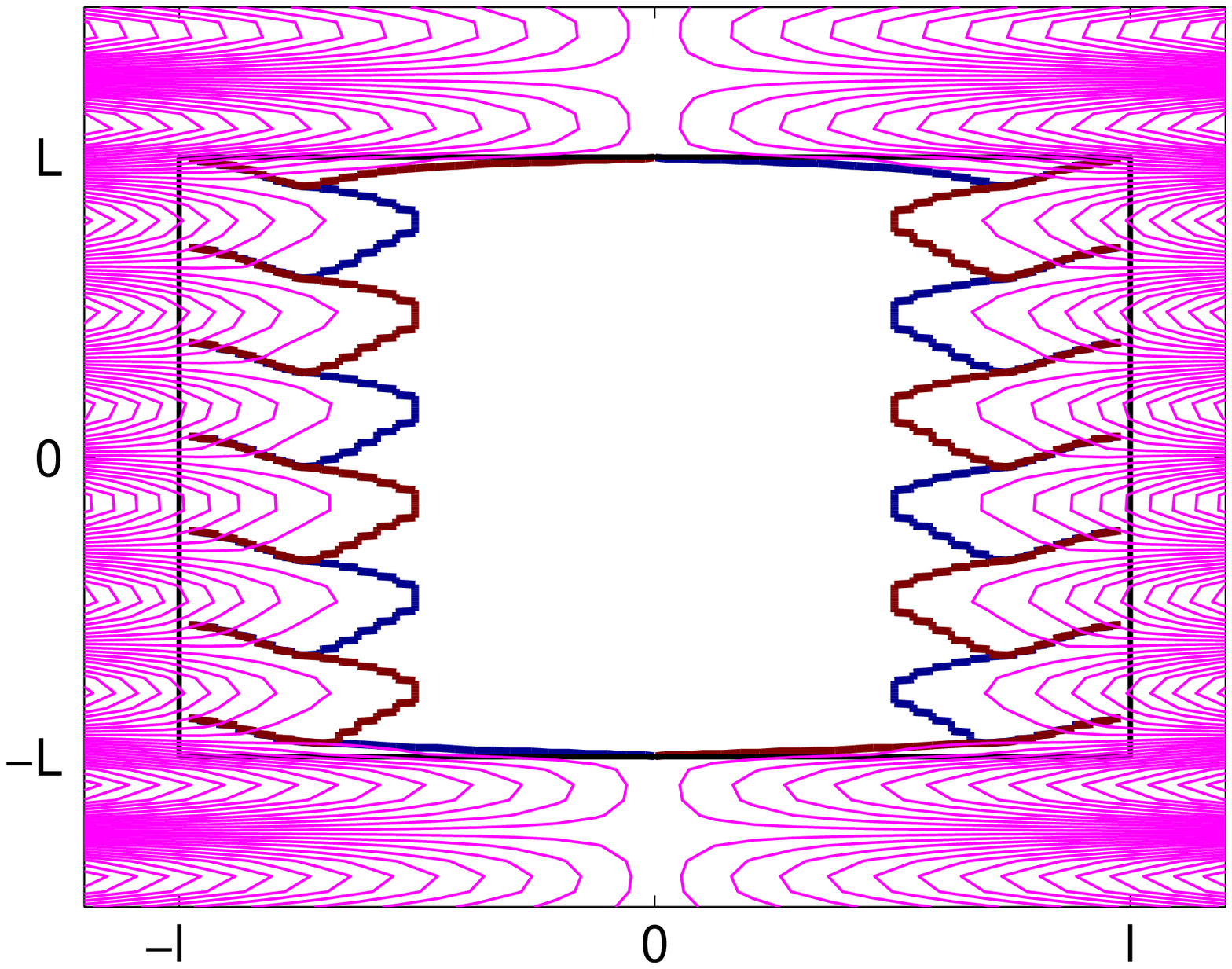}\hspace{.1cm}
\includegraphics[width=3.cm,height=8cm]{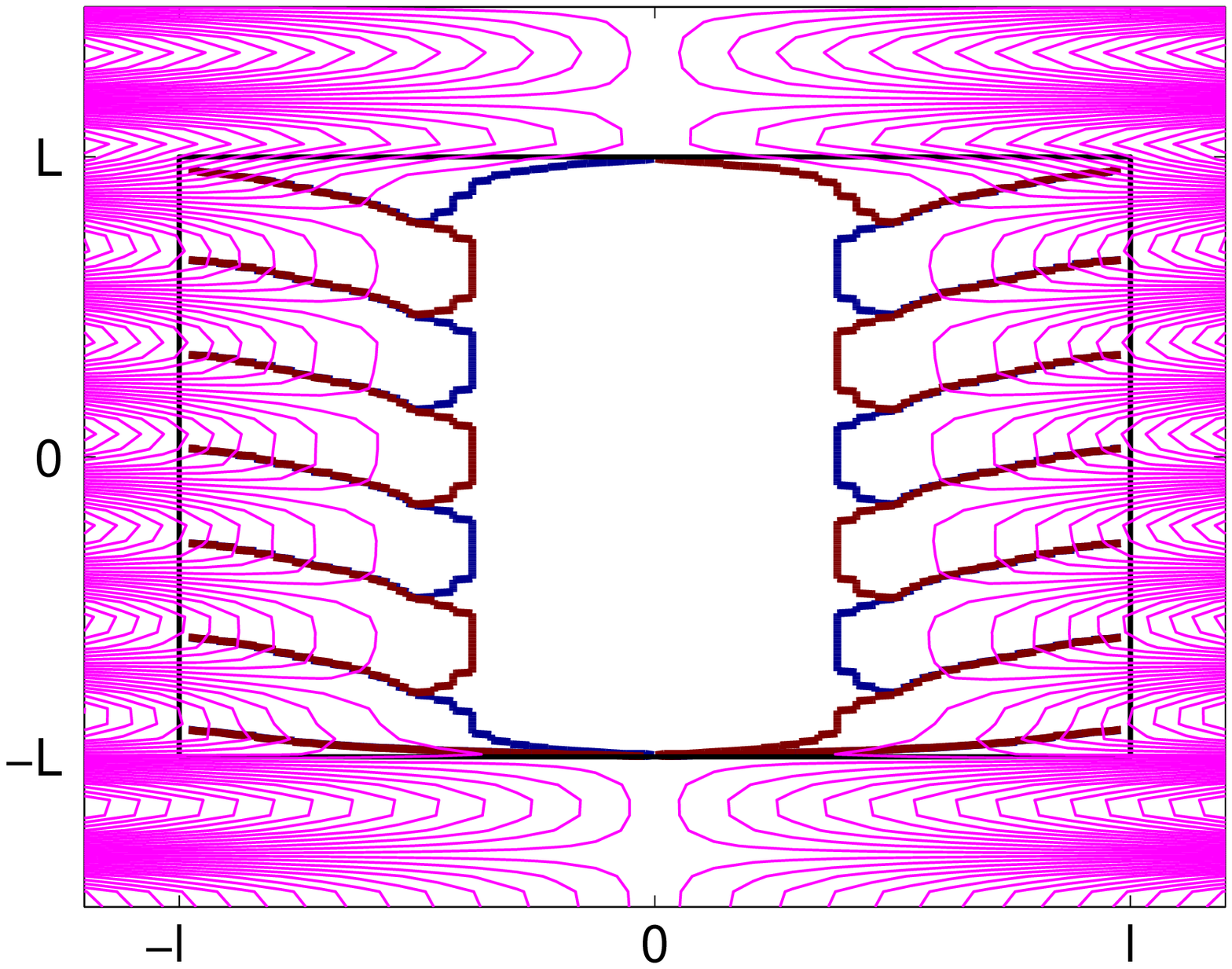}\hspace{.1cm}
\includegraphics[width=3.cm,height=8cm]{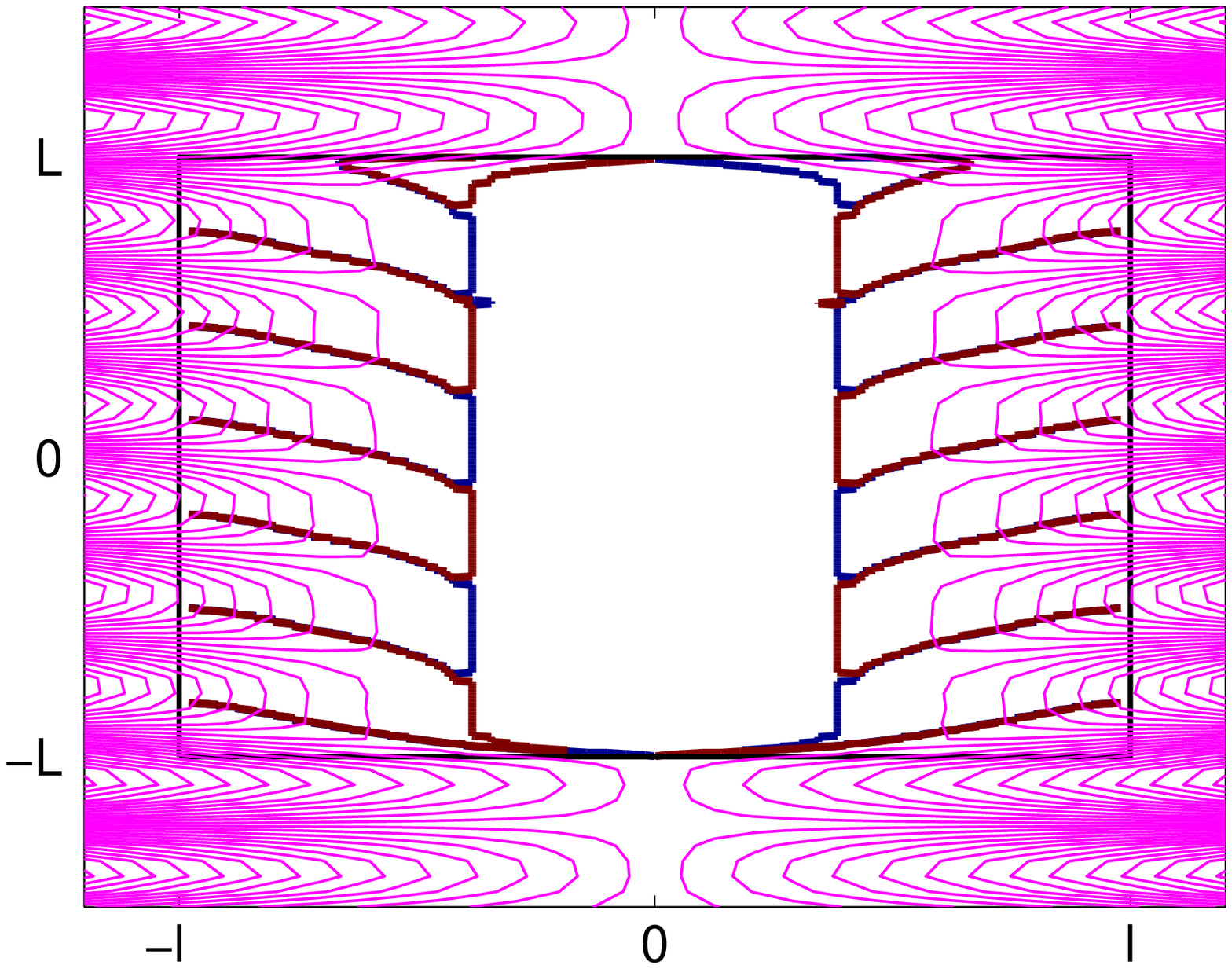} \caption{Running wave.
Penetration of the magnetic field induced by an external current
$\mathbf{J}_e=i_{e}(t) \sin \{
k(y-vt)\}[\delta(x-a)-\delta(x+a)]\mathbf{e}_z$ into a cylindrical
superconductor with the cross section $\Omega=\{-l\leq x\leq l,\
-L\leq y\leq L\}$. The wave amplitude $i_{e}$ grows linearly from
zero at $t=0$ to its maximal value $i_{e0}$ at $t=t_0$ and then
remains constant. The physical parameters: $l=1/k,\ L=10/k,\
a=1.1/k,\ t_0=4/kv,\ i_{e0}=J_c/k$. Shown for $t=[0.8,\ 1.6,\
3.2,\ 5.6,\ 16.0]/kv$ are: the boundaries of plus- and minus
critical current zones, the magnetic field lines, and the cylinder
cross section. The parameters of numerical scheme: $50\times 500$
regular mesh of piecewise constant finite elements, time step
$0.8/kv$.} \label{Fig1} \end{figure}

In our first example (Fig. \ref{Fig1}) the external current is
given as a running wave, $J_{e}=i_{e}(t) \sin \{
k(y-vt)\}[\delta(x-a)-\delta(x+a)],$ with the amplitude $i_{e}(t)$
growing from zero to its maximal value and then remaining
constant.  The magnetic field penetrates from the surface of the
superconductor where  alternating domains of plus- and minus
critical current densities appear and start to follow the wave.
The shape of these domains stabilizes and, after an initial
transient period, they completely occupy a near-surface zone of a
constant depth and move through this zone with the wave velocity
$v$.

\begin{figure}[th]
\centering
\includegraphics[width=3.cm,height=8cm]{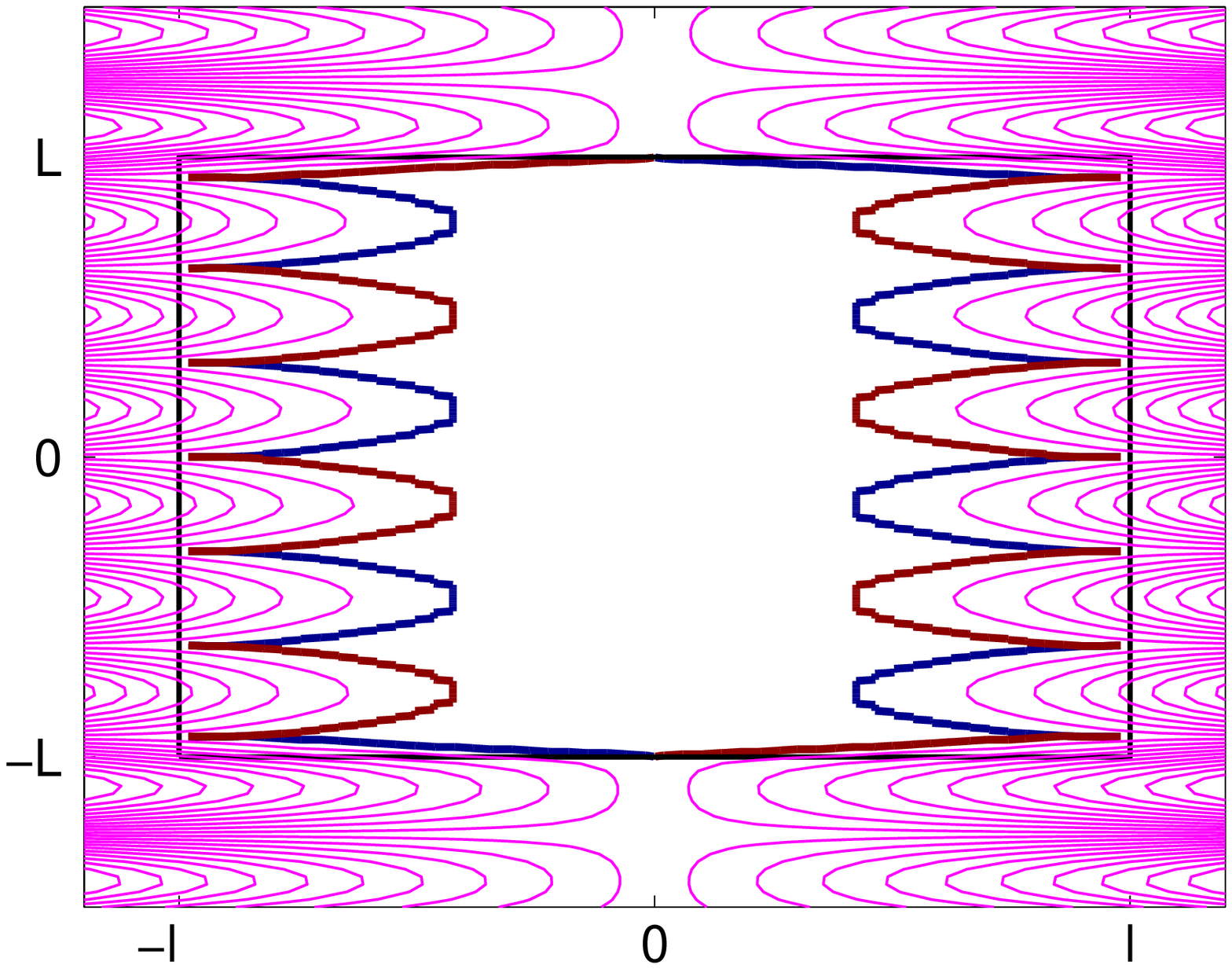}\hspace{.1cm}
\includegraphics[width=3.cm,height=8cm]{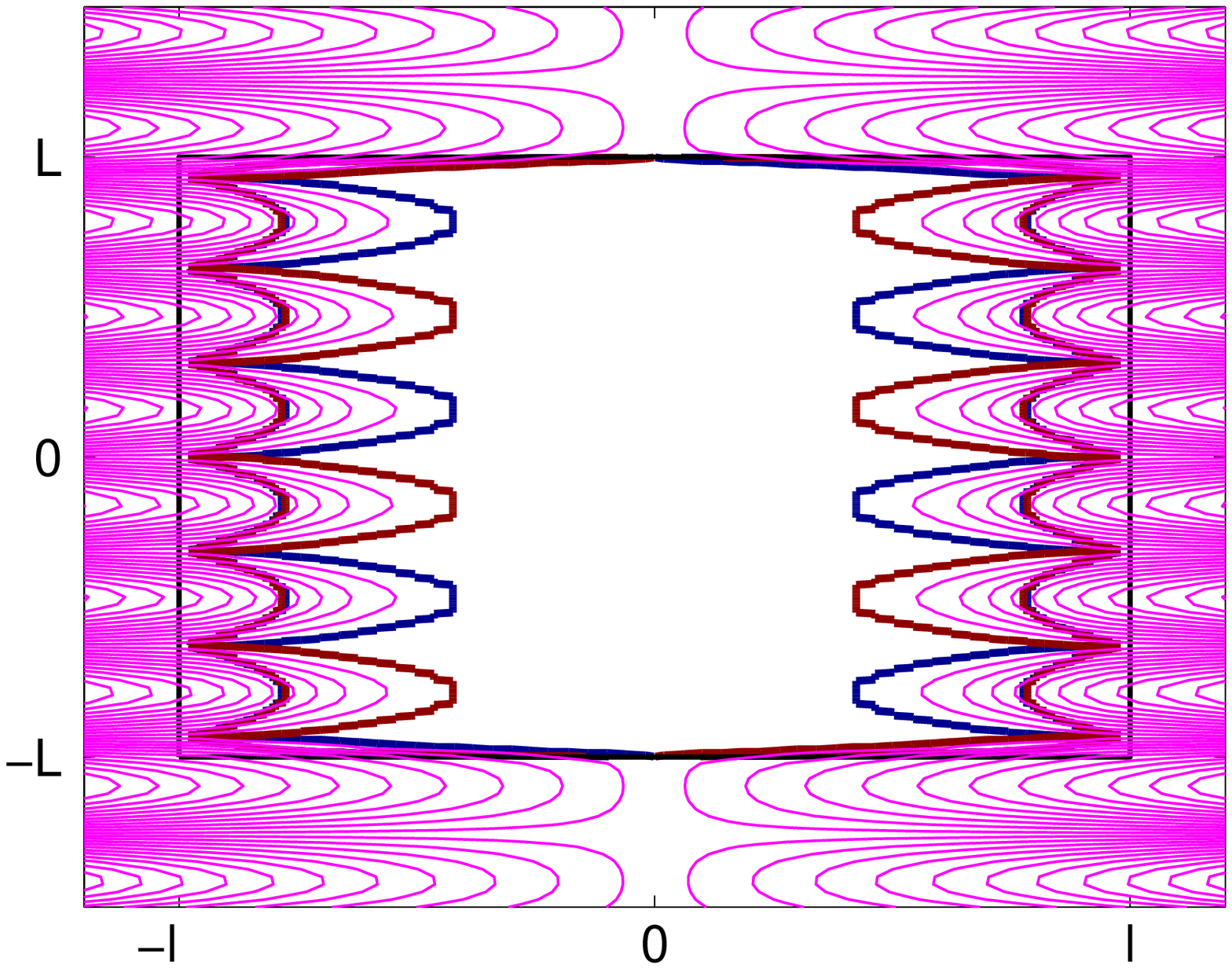}\hspace{.1cm}
\includegraphics[width=3.cm,height=8cm]{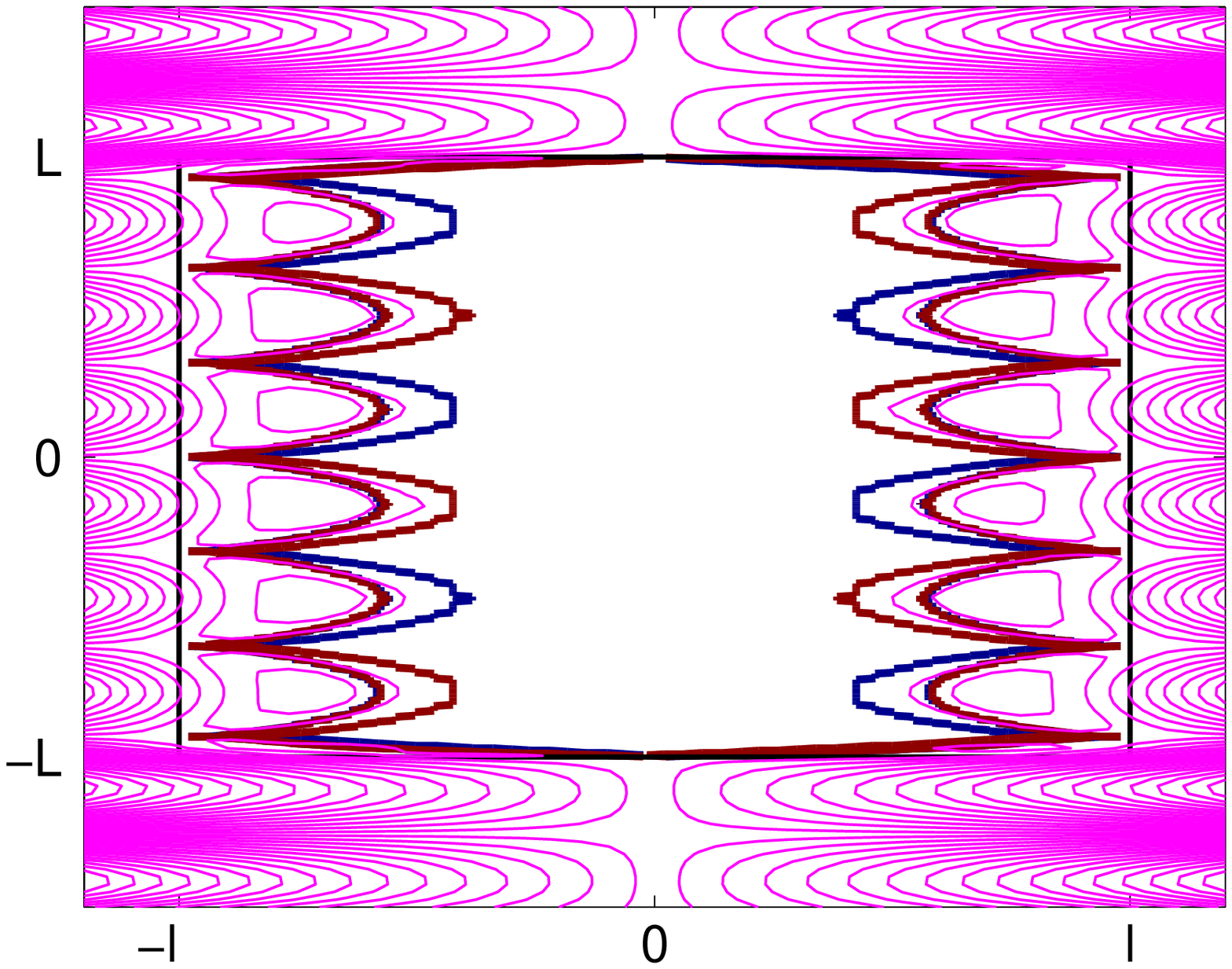}\hspace{.1cm}
\includegraphics[width=3.cm,height=8cm]{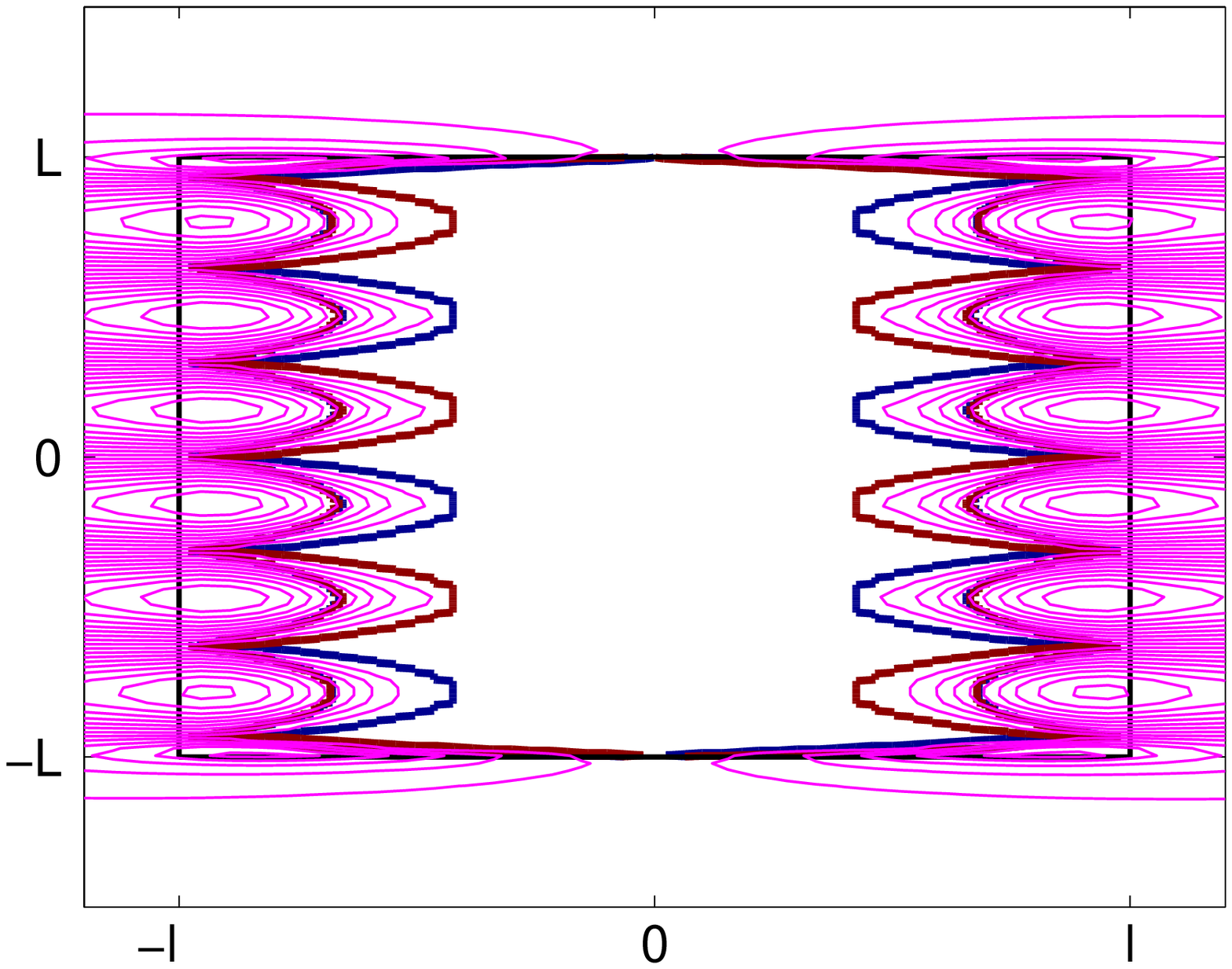}\hspace{.1cm}
\includegraphics[width=3.cm,height=8cm]{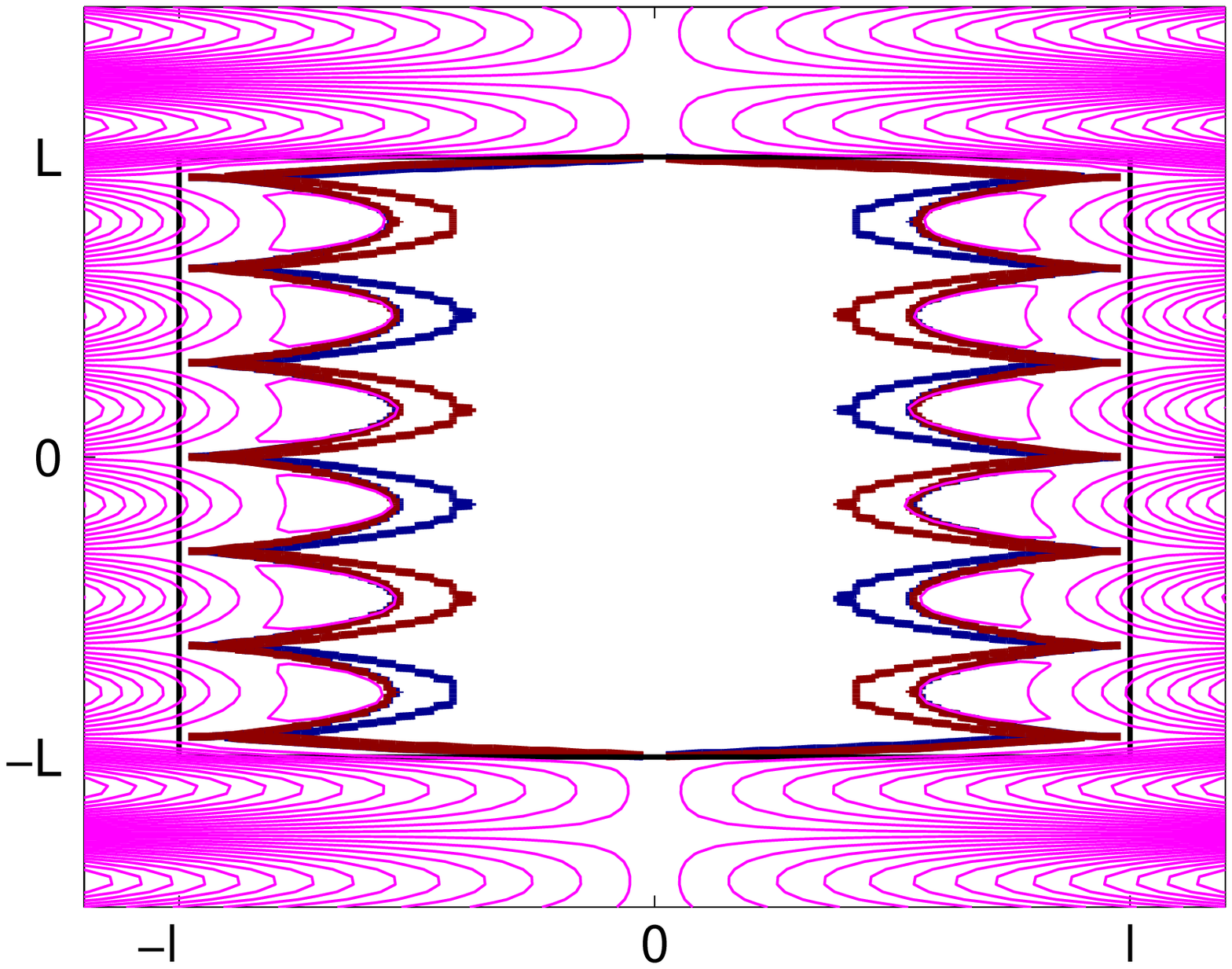}
\caption{Standing wave. The external magnetic field is induced by
the current   $\mathbf{J}_e=i_{e} \sin (ky)\sin(2\pi f t
)[\delta(x-a)-\delta(x+a)]\mathbf{e}_z$.  Here $l=1/k,\ L=10/k,\
a=1.1/k,\ i_{e}=J_c/k$. Shown for $t=[0.24,\ 0.44,\ 0.56,\ 1.00,\
1.12]/f$ are: the boundaries of plus- and minus critical current
zones, the magnetic field lines, and the cylinder cross section.
Regular finite element mesh $50\times 500$, time step $0.04/f$.}
\label{Fig2}
\end{figure}

The second example (Fig. \ref{Fig2}) illustrates another typical
situation: here the magnetic field fluctuations are induced by the
external current in the form of a standing wave, $J_{e}=i_{e} \sin
(ky)\sin(2\pi f t)[\delta(x-a)-\delta(x+a)]$. At $t=0$ domains of
plus- and minus critical current densities, shielding the external
magnetic field, appear at the superconductor surface  and start to
propagate inside. When the external field reaches its maximal
strength at $t={T}/{4}$, the propagation stops (here $T=1/f$). As
the field becomes weaker, the boundary of the critical current
regions does not, however, retreat. Instead, similarly to the case
of an alternating uniform external field, to compensate the
decreasing external field there appear surface domains of the
opposite critical current densities. These domains propagate
inside, sweep out the previous ones at
$t={3T}/{4}$, and the process becomes periodic. 
\section{Asymptotic solution\label{Asimp_sec}}
Let an infinite slab $-l\leq x\leq l$ be placed into a magnetic
field produced by the sheet current (\ref{JeV}) or (\ref{JeS}). We
want to find the established  periodic in time distribution of
induced current density. It follows from (\ref{vi1}) that the
time-periodic part of current density we are interested in does
not depend on the permanent part of external current. We set
$I_e=0$ again just to simplify the consideration; this is not a
limitation of the method employed.

It is not difficult to find  a distribution of surface current
density, $i_s(y,t)$, such that the current
$i_s(y,t)[\delta(x+l)-\delta(x-l)]$  shields the superconductor
from the external field. Clearly, complete shielding occurs if the
magnetic vector potential of this current, $A_s$, compensates the
external magnetic potential inside the superconductor,
$$A_e+A_s=0$$
for $-l<x<l$. Using expressions (\ref{AeV}) and (\ref{AeS}), we
find that the shielding would be achieved if we set
\begin{eqnarray}i_{s}=-e^{-k(a-l)} i_{e} \sin
\{
k(y-vt)\},\label{IsV}\\
i_{s}=-e^{-k(a-l)} i_{e} \sin (ky)\sin(2\pi f
t)\label{IsS}\end{eqnarray} for the running and standing waves,
(\ref{JeV}) and (\ref{JeS}), respectively.

We now assume that the depth $\Delta$ to which  fluctuations
penetrate into the superconductor is much smaller than the
fluctuation wavelength, $k\Delta \ll 1,$ interpret the shielding
surface currents as the integrals of bulk current densities across
a narrow penetration zone, and  find the asymptotic distribution
of bulk current density inside this zone analytically. First, let
us note that for each $y$ the surface current $i_s(y,t)$ reaches
its extremal values when the whole penetration zone (see Figs.
\ref{Fig1} and \ref{Fig2}) is occupied by the critical current
density of the same sign. Therefore, the penetration depth can be
calculated as $\Delta=\max(i_s)/J_c$, which gives
$\Delta=e^{-k(a-l)} i_{e}/J_c$ for the running wave and
$\Delta(y)=e^{-k(a-l)} i_{e}|\sin(ky)|/J_c$ for the standing wave.
We see that for both wave types the fluctuations may be regarded
small if
\begin{equation}\nu=k\Delta_0\ll 1,\label{cond2}\end{equation} where $\Delta_0=
i_{e}e^{-k(a-l)}/J_c$.

If, at time $t$, the current $i_s(y,t)$ is neither maximal nor
minimal, the penetration zone $-l\le x\le -l+\Delta$ (where
$\Delta=\Delta_0$  for the running wave and
$\Delta=\Delta_0|\sin(ky)|$ for the standing wave) contains
regions of plus- and minus critical current densities. The
near-surface region $-l<x<-l+\sigma(y,t)$ appears at the time when
$\partial_ti_s(y,t)=0$ and, as it propagates inside, the current
density there is $J_c\mbox{sign}\{\partial_ti_s(y,t)\}$. The rest
of the penetration zone, $-l+\sigma(y,t)<x<-l+\Delta$, is occupied
by the critical current density of the opposite sign.  Comparing
$i_s$ with the integral of current density across the penetration
zone, we find the moving boundary $\sigma(y,t)$: \begin{equation}
\sigma(y,t)=\frac{1}{2}\left(\Delta+\frac{i_s(y,t)}{J_c}\mbox{sign}\{\partial_t
i_s(y,t)\}\right).\label{S}\end{equation}   Near the slab surface
$x=+l$ the current density distribution is antisymmetric.

Simple physical arguments were used above to obtain the asymptotic
solutions for  weak penetration: postulating the solution
structure, we spread the shielding surface current into a bulk
current in the near-surface zone. In a similar way,  weak
penetration of alternating uniform field into a perpendicular
circular cylinder has been studied in \cite{Carr,Ashkin}. Although
the chosen surface current would have shield the superconductor
from the external magnetic field completely, spreading this
current into the bulk makes shielding imperfect. As will be shown
below, the remaining field is of the order $O(\nu)$. We will now
extend our arguments and present the obtained asymptotic solutions
as zero order terms of consistent asymptotic expansions.
\subsection{Running wave} It is not
difficult to see that the asymptotic distribution of current
density inside the penetration zone $-l\le x\le -l+\Delta_0$,
obtained for $\nu=k\Delta_0\ll 1$, can be presented as
\begin{equation}J=-J_cS(k(y-vt)-\Psi_0(\varsigma/\Delta_0)),\ \ \
\varsigma=x+l\in [0,\Delta_0],\label{J0}\end{equation} where
$S(z)=\mbox{sign}(\sin(z))$ is a $2\pi$-periodic step-function and
$\Psi_0(u)=\arcsin(1-2u)$.

Let us now look for the current density
\begin{equation} J=-J_cS(k(y-vt)-\Psi(\varsigma/\Delta)),\ \ \
\varsigma=x+l\in [0,\Delta]\label{Jmod}\end{equation} where
\begin{eqnarray}\Delta=\Delta_0(1+r_1\nu+r_2\nu^2+...),\label{DltaExp}\\
\Psi(u)=\Psi_0(u)+\nu\Psi_1(u)+\nu^2\Psi_2(u)+...\label{PsiExp}\end{eqnarray}
are such that the current (\ref{Jmod}), jointly with the opposite
one near another superconductor surface, shield the  external
magnetic field.

We partly solve this problem in Appendix I by showing first that
the vector potential produced by the zero-order approximation
(\ref{J0}) to current density  compensates the potential of
external current inside the superconductor (but outside the
penetration zone) up to the $O(\nu)$ terms; hence the external
magnetic field inside the superconductor is shielded up to the
same order in $\nu$. We find then the first order corrections,
$r_1$ and $\Psi_1$, ensuring shielding of the magnetic field up to
the second order, $O(\nu^2)$.

\begin{figure}[h]
\centering
\includegraphics[width=5cm,height=12cm]{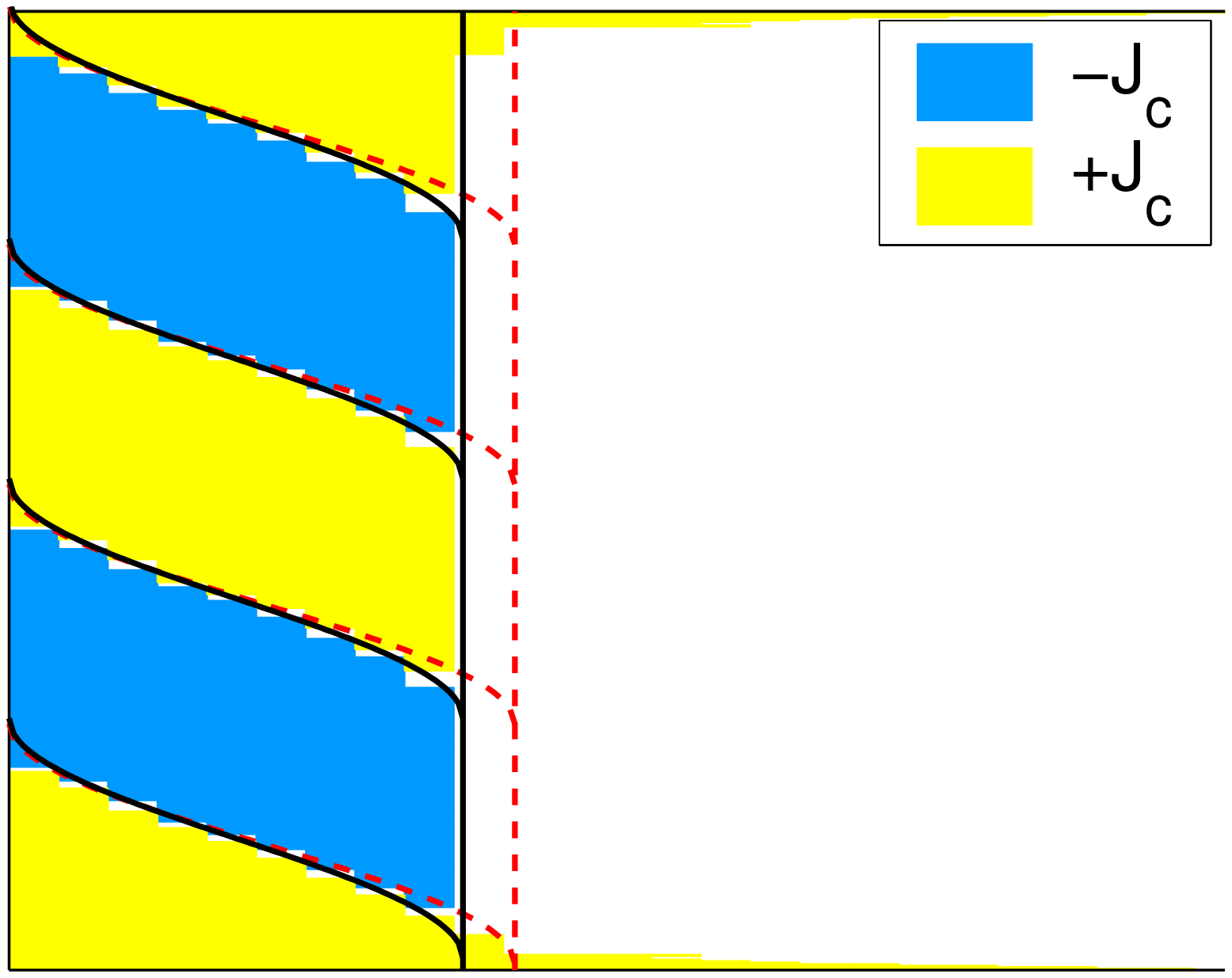}\hspace{2cm}
\includegraphics[width=5cm,height=12cm]{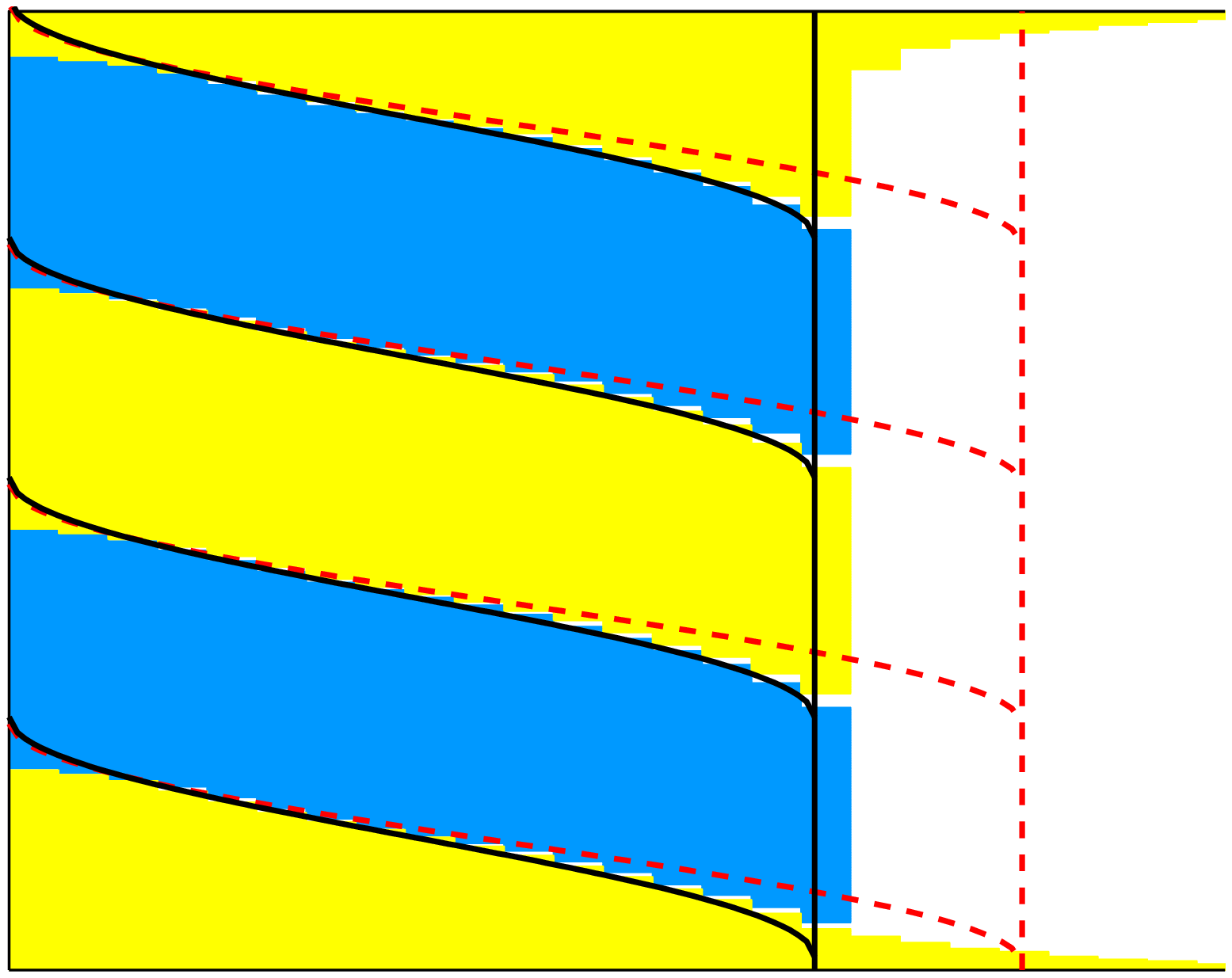}
\caption{Penetration of the running wave fluctuations. Domains of
plus- and minus critical current densities (numerical solution)
and their asymptotic boundaries: "- -" -- zero order
approximation, "---" -- first order approximation. Left: $i_{e}
=0.25J_c/k$ and $\nu=0.2$; right: $i_{e} =0.5J_c/k$ and $\nu=0.4$.
In both cases, $l=0.5/k,\ a=0.7/k,\ L=2\pi/k$. Different scales in
x and y.} \label{FigAs2RW}
\end{figure}

It turns out (see Appendix I) that a nonsingular function $\Psi_1$
can be found only if $r_1=-\frac{1}{2}$, so the penetration depth
becomes
$$\Delta=\Delta_0(1-\frac{1}{2}\nu+O(\nu^2)).$$
It is further shown that the external magnetic field is shielded
up to the second order in $\nu$ if
\begin{equation}\Psi_1(\varsigma/\Delta)=\sum_{m=0}^{M}a_mU_{2m}(1-2\varsigma/\Delta),\label{Psi1_mod}
\end{equation}
where $U_n(\tau)=\sin((n+1)\arccos(\tau))/\sin(\arccos(\tau))$ are
the Chebyshev polynomials of the second kind,
\begin{equation}
a_m=-\frac{1}{\pi}\left(\frac{1}{4(m+1)^2-1}+\frac{1}{4m^2-1}\right),\label{am}\end{equation}
and $M$ satisfies the conditions $$\nu\ln(M+1)\ll \pi$$ if $M>1$
(or $\nu\ll\pi$ if $M=0,1$) and
$$\frac{e^{-(2M+3)\nu}}{2M+1}\lesssim O(\nu).$$
The first of these conditions ensures that $|\nu \Psi_1|\ll 1$,
the second provides for the $O(\nu^2)$ shielding of magnetic
field. The conditions are easy to satisfy and, usually, only a few
terms of the series are needed. Thus, in the examples in Fig.
\ref{FigAs2RW}, we used $M=1$ for $\nu=0.2$ and $M=0$ for
$\nu=0.4$.

\subsection{Standing wave} We find first a correction to
the asymptotic penetration depth $\Delta(y)=\Delta_0|\sin(ky)|$.
Let us choose a moment when the external field is the strongest,
e.g., $t=T/4$, so that for each $y$ the induced current density is
\begin{equation} J=-J_cS(ky),\label{JSt}\end{equation} in the
whole penetration zone $0\le \varsigma=x+l\le \Delta(y)$ (and is
opposite in the zone near another slab surface). We will now
assume that
\begin{equation}\Delta(y)=
\Delta_0(|\sin(ky)|+\nu\Psi_1(ky)+\nu^2\Psi_2(ky)+...)\label{PsiSt}
\end{equation}
and find a correction $\Psi_1(s)$ that ensures better shielding of
the external  field for this moment of time.

\begin{figure}[h]
\centering
\includegraphics[width=5cm,height=12cm]{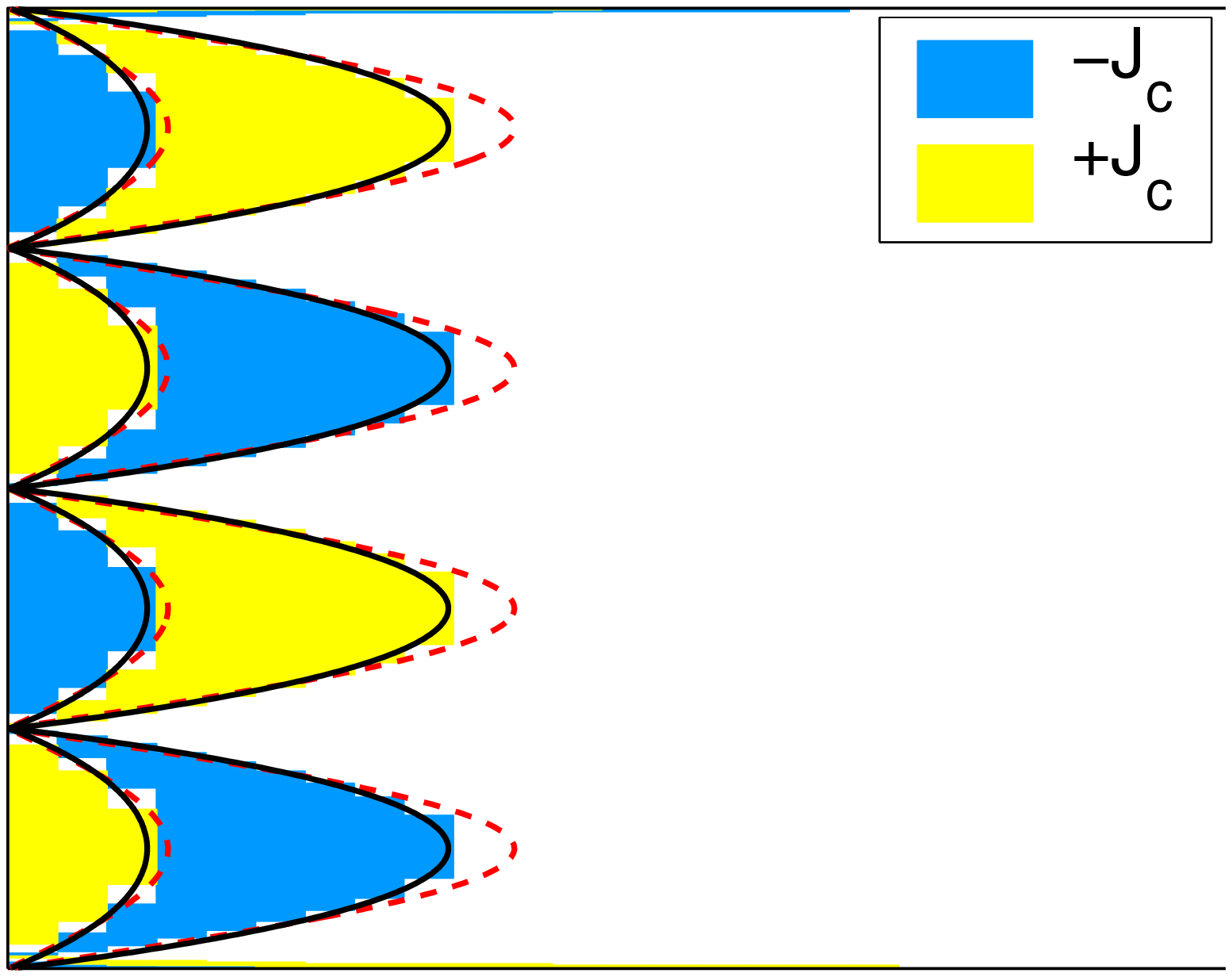}\hspace{2cm}
\includegraphics[width=5cm,height=12cm]{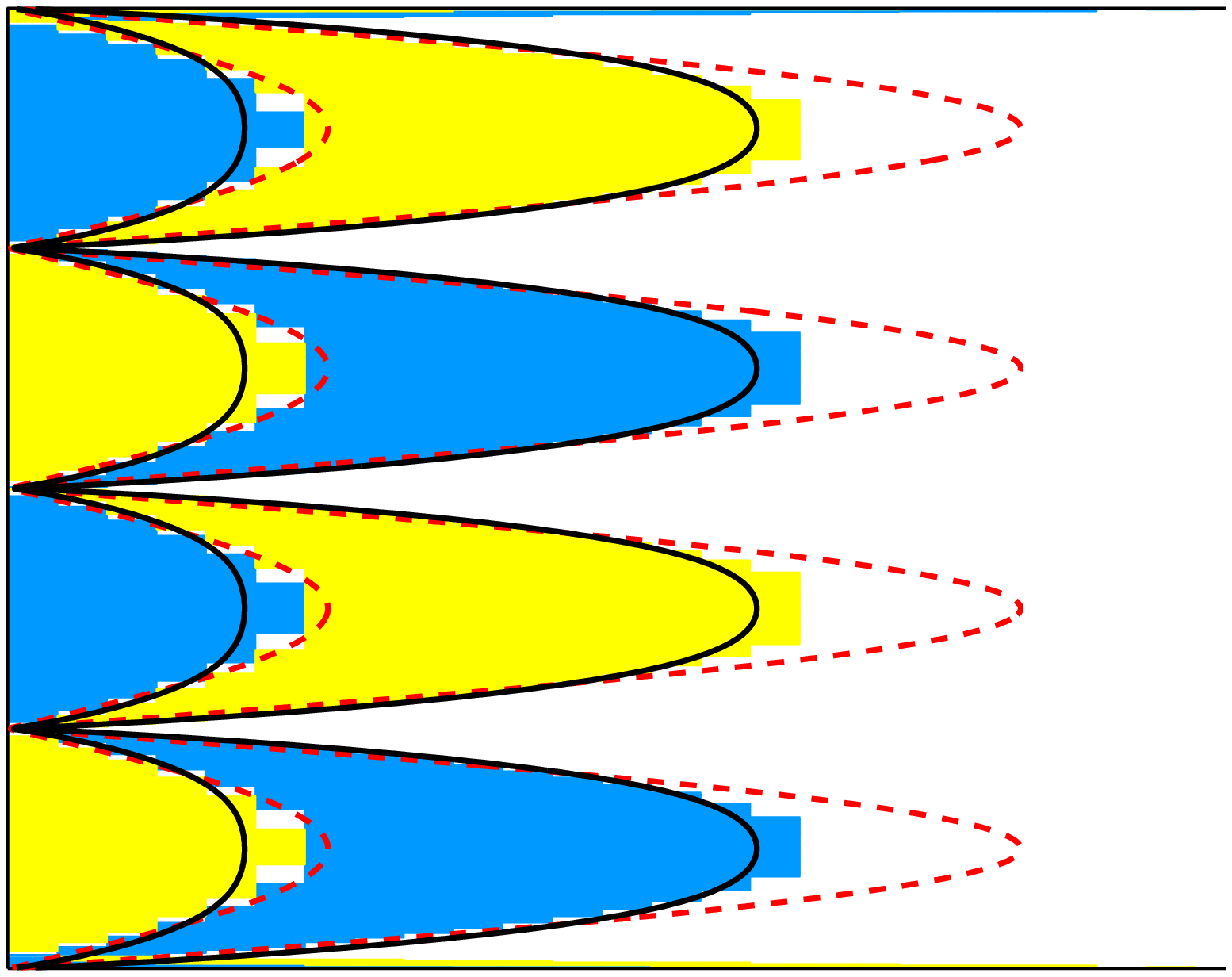}
\caption{Penetration of the standing wave fluctuations. Notations
and parameters as in the previous figure.} \label{FigAs2SW}
\end{figure}

It can be shown (see Appendix II) that the magnetic field is
shielded up to the second order in $\nu$ if
\begin{equation}\Psi_1(s)=
-\frac{1}{\pi}\mbox{sign}(\sin(s))[2\sin(s)+\sin(2s)
\ln(|\tan(s/2)|)].\label{Psi1St}\end{equation}
Equations (\ref{JSt}), (\ref{PsiSt}), and (\ref{Psi1St}) give the
asymptotic distribution of current for $t=T/4$. It is now easy to
obtain the solution for any time moment. Let, for example,
$T/4<t<3T/4$. Then the closest to the surface $x=-l$ part of the
penetration zone is occupied by the current density $+J_cS(ky)$.
We can present this as a superposition of the current density
$-J_cS(ky)$ in the whole penetration zone and the current
$+2J_cS(ky)$ in the part of this zone near the surface. Up to the
second order terms, the current $-J_cS(ky)$ shields the external
current $i_e\sin(ky)[\delta(x+a)-\delta(x-a)]$, so the opposite
current density $+2J_cS(ky)$ has to shield the external current
$i_e(\sin(2\pi ft)-1)\sin(ky)[\delta(x+a)-\delta(x-a)]$.  This
means that the boundary between the two zones must be
\begin{equation}\widetilde{\Delta}(y,t)=\frac{\Delta(y)}{2}(1-\sin(2\pi f
t)).\label{D_tilde}\end{equation} In Fig. \ref{FigAs2SW}, we
compare the numerical and asymptotic solutions for two values of
parameter $\nu$. As in the running wave case, these solutions are
close.
\section{AC losses\label{ACloss}}
Let the external magnetic field be periodic in time, e.g., induced
by a periodic current density $\mathbf{J}_e(r,t)$, and $T$ its
period. Then there establishes a periodic induced current density
in the superconductor, $\mathbf{J}(r,t)$. Suppose this latter
function was found and it is needed to calculate the energy
losses,
$$P=\frac{1}{T}\int_0^T\int_{\Omega}\mathbf{J}\cdot \mathbf{E}\,d\Omega dt=
\frac{1}{T}\int_0^T\int_{\Omega}\rho \mathbf{J}^2\,d\Omega dt.$$
This would be an easy task for a usual conductor with the known
resistivity. However, for type-II superconductors $\rho(r,t)$ is
an effective resistivity caused by the movement of magnetic
vortices and is not known \emph{a priory}. Mathematically, in the
Bean model, this is a dual variable excluded by transition to the
variational formulation; accurate calculation of $\rho$ (or
$\mathbf{E}$) in the general case is difficult. To avoid this
complication, we employ a  method based on the magnetic potential
representation of electric field (\ref{E}) and similar to those
suggested in \cite{Ashkin,Bossavit1} but applicable to nonuniform
external fields.

Since $\mathbf{J}$ satisfies the zero divergence conditions
(\ref{div0}), for any gauge holds
$$P=\frac{1}{T}\int_0^T\int_{\Omega}\mathbf{J}\cdot \mathbf{E}\,d\Omega dt
=-\frac{1}{T} \int_0^T\int_{\Omega}\mathbf{J}\cdot
\partial_t\mathbf{A}\,d\Omega dt=-\frac{\mu_0}{T}
\int_0^T\int_{\Omega}\mathbf{J}\cdot
G*\partial_t\mathbf{J}\,d\Omega dt-\frac{1}{T}
\int_0^T\int_{\Omega}\mathbf{J}\cdot
\partial_t\mathbf{A}_e\,d\Omega dt.$$ It is easy
to see that
$$\int_0^T\int_{\Omega}\mathbf{J}\cdot G*\partial_t\mathbf{J}\,d\Omega dt=\int_0^T
\frac{d}{dt}\left\{\frac{1}{2}\int_{\Omega}\mathbf{J}\cdot
G*\mathbf{J}\,d\Omega\right\} dt=0$$ due to periodicity of
$\mathbf{J}$. Hence we obtain
$$P=
-\frac{1}{T}\int_0^T\int_{\Omega}\mathbf{J}\cdot\partial_t\mathbf{A}_e\,d\Omega
dt$$  and, since the time period of the product
$\mathbf{J}\cdot\partial_t\mathbf{A}_e$ is
 $T/2$, we can also write
\begin{equation}
P=-\frac{2}{T}\int_{t_0}^{t_0+T/2}\int_{\Omega}\mathbf{J}\cdot\partial_t\mathbf{A}_e\,d\Omega
dt\label{P}\end{equation} for arbitrary time moment $t_0$.
\subsection{AC losses for running wave}
Clearly, only half of the slab may be considered due to symmetry
and we can now use (\ref{P}) with $\Omega=(-l,0)$ and $T=2\pi/kv$
to find the asymptotic AC losses for small fluctuations.
Obviously, the value of integral (\ref{P}) must be the same for
all $y$ and equal to the rate of AC losses per unit of slab
surface. We neglect the second order terms in (\ref{DltaExp}),
(\ref{PsiExp}) and assume
\begin{equation}\Psi=\Psi_0+\nu\Psi_1,\ \ \
\Delta=\Delta_0\left(1-\frac{\nu}{2}\right).\label{PsDe}\end{equation}
To simplify computations, let us take $y=0$, so that (\ref{Jmod})
gives \[ J|_{y=0}=J_cS(kvt+\Psi(\varsigma/\Delta)),\ \ \
\varsigma=x+l\in[0,\Delta],\] and choose $t_0=-\Psi(0)/kv$. For
small $\nu$ the function $\Psi(\varsigma/\Delta)$ is close to
$\Psi_0(\varsigma/\Delta)$ and monotonically decreases for
$\varsigma\in[0,\Delta]$. Therefore,
$kvt_0+\Psi(\varsigma/\Delta)$ changes monotonically from zero at
$\varsigma=0$ to $kvt_0+\Psi(1)=-\pi$ at $\varsigma=\Delta$.
Indeed, since $\Psi_1(0)=\Psi_1(1)$, we have
\begin{equation}\Psi(1)=\Psi_0(1)+\nu\Psi_1(1)=-\pi+\Psi_0(0)+\nu\Psi_1(0)=-\pi+\Psi(0).
\label{pi}\end{equation} Hence, for $y=0$ the whole penetration
zone at $t=t_0$ is occupied by the current density $-J_c$. For
$t_0<t<t_0+T/2$ there is a $+J_c$-current-density zone propagating
inside and sweeping out the $-J_c$ zone at $t=t_0+T/2$. The moving
boundary between the two zones is determined by the condition
$kvt+\Psi(\varsigma/\Delta)=0$. Taking this into account we
rewrite (\ref{P}) as
\[P=-\frac{2J_c}{T}\int_0^{\Delta}\left.\left\{-\int_{-\Psi(0)/kv}^{-\Psi(\varsigma/\Delta)/kv}
\partial_tA_edt+\int_{-\Psi(\varsigma/\Delta)/kv}^{-\Psi(1)/kv}\partial_tA_edt\right\}
\right|_{\tiny
\begin{array}{l}x=-l+\varsigma\\y=0\end{array}}d\varsigma=\]
\[
-\frac{2\mu_0i_eJ_c\Delta}{kT}e^{-ka}\int_{0}^{1}[2\sin\Psi(u)-
\sin\Psi(1)-\sin\Psi(0)]\sinh\left(-kl+\nu
\frac{\Delta}{\Delta_0}u\right)du,\] since $A_e(x,y,t)$ is given
by (\ref{AeV}). Note that $\sin\Psi(1)+\sin\Psi(0)=0$ because of
(\ref{pi}); also $i_ee^{-ka}=J_c\Delta_0e^{-kl}$. Substituting
$\Delta$ and $\Psi$ from (\ref{PsDe}) we obtain
\[P=\frac{4\mu_0J_c^2\Delta_0}{k^2T}P_0,\]
where, up to the higher order terms,
\begin{equation}\begin{split}P_0& =-\nu
e^{-kl}\left(1-\frac{1}{2}\nu\right)
\int_{0}^{1}\sin\{\arcsin(1-2u)+\nu\Psi_1(u)\}\sinh(-kl+\nu u)du\\
& =-\frac{\nu^2}{2}e^{-kl}\left[-\frac{1}{3}\cosh(-kl)+
\sinh(-kl)\int_{-1}^1\Psi_1\left(\frac{1-\tau}{2}\right)\sqrt{1-\tau^2}d\tau\right]\end{split}\label{P0R}\end{equation}
Here $\tau=1-2u$ and $\Psi_1$ is given by (\ref{Psi1_mod}) as a
finite series of the Chebyshev polynomials. We see that the
leading AC loss term depends on
 the first order correction to solution (\ref{J0}).
Since
\[\int_{-1}^1
U_{2m}(\tau)\sqrt{1-\tau^2}d\tau=\left\{\begin{array}{ll}\frac{\pi}{2}\ &\ m=0,\\
0 &\ m>0\end{array}\right.\] only the first term of the series
(\ref{Psi1_mod}) gives input into AC losses. We find
\[P_0=\frac{\nu^2}{2}e^{-kl}\left[\frac{1}{3}\cosh(-kl)-
\frac{a_0\pi}{2}\sinh(-kl)\right]=\frac{\nu^2}{6}\] and, finally,
the asymptotic losses for weak penetration,
\begin{equation}
P=\frac{2\mu_0J_c^2\Delta_0}{3k^2T}\nu^2=
\frac{2}{3}f\mu_0J_c^2\Delta_0^3,\label{P_RW}\end{equation} where
$f=1/T$ is the frequency of fluctuations. Note that the expression
obtained would coincide with the well known formula for hysteresis
losses caused by  fluctuations of spatially uniform external
magnetic field (incomplete penetration),
$P=\frac{2}{3}f\mu_0H_0^3/J_c$, if we rewrite this latter formula
not for the fluctuation amplitude $H_0$ but using the penetration
depth $\Delta_0$, equal to $H_0/J_c$ for the uniform field.
\subsection{AC losses for standing wave}
To estimate the average asymptotic AC losses per unit of slab
surface we modify slightly  the formula (\ref{P}) to calculate the
density of these losses,
\[P(y)=-\frac{2}{T}\int_{T/4}^{3T/4}\int_{0}^{\Delta(y)}\partial_tA_eJ
d\varsigma\,dt,\] where $T=1/f$,  then average over half the
fluctuation wavelength,
\[\langle P\rangle=\frac{2}{\lambda}\int_{0}^{\lambda/2}P(y)dy,\]
where $\lambda=2\pi/k$. Assuming $T/4<t<3T/4$ and $0<y<\pi/k$ we
get $J=+J_c$  for $0<\varsigma<\widetilde{\Delta}(y,t)$ and
$J=-J_c$ for $\widetilde{\Delta}(y,t)<\varsigma<\Delta(y)$, where
$\Delta(y)=\Delta_0(|\sin(ky)|+\nu\Psi_1(ky)+O(\nu^2))$ and
$\widetilde{\Delta}(y,t)$ is given by (\ref{D_tilde}).
Substituting $I_e=0$, $x=-l+\varsigma$ into the vector potential
(\ref{AeS}) and integrating, we obtain
\[P(y)=-\frac{2J_c}{T}\int_{T/4}^{3T/4}
\left\{\int_0^{\widetilde{\Delta}(y,t)}\partial_tA_ed\varsigma-
\int_{\widetilde{\Delta}(y,t)}^{\Delta(y)}\partial_tA_ed\varsigma\right\}dt
=\frac{4}{k^2}\mu_0fJ_c^2\Delta_0P_0(y),\] where
\begin{equation}P_0(y)=e^{-kl}\sin(ky)\left[\cosh(-kl)+\cosh\left(-kl+\nu\frac{\Delta(y)}{\Delta_0}\right)-
\frac{\sinh\left(-kl+\nu\frac{\Delta(y)}{\Delta_0}\right)-
\sinh(-kl)}{\frac{\nu\Delta(y)}{2\Delta_0}}\right].\label{P0S}\end{equation}
For $0\le y\le\pi/k$ we have  $\Delta(y)/\Delta_0 =\sin(ky)+\nu
\Psi_1(ky)+O(\nu^2)$ and, up to the higher order terms,
\[P_0(y)=e^{-kl}\left[\frac{\nu^2}{6}\sin^3(ky)\cosh(kl)+\nu^3
\left(\frac{1}{3}\Psi_1(ky)\sin^2(ky)\cosh(kl)
-\frac{1}{12}\sin^4(ky)\sinh(kl)\right)\right].\] Except for the
points very close to $y=0$ or $\pi/k$, where the term containing
$\Psi_1(ky)$ dominates but the losses are negligible,
\begin{equation}P(y)\approx
\frac{2}{3k^2}\mu_0fJ_c^2\Delta_0\nu^2\sin^3(ky)e^{-kl}\cosh(kl)=
\frac{2}{3}\mu_0fJ_c^2\Delta_0^3(y)\sin^3(ky)\frac{1+e^{-2kl}}{2}.\label{PlocStW}\end{equation}
This is close to the density of losses in an alternating uniform
field if $kl\ll 1$ and is twice smaller if $kl\gg 1$, provided the
penetration depth is the same. Finally, we find the asymptotic
average AC losses per unit of slab surface for the standing wave
fluctuations:
\begin{equation}\langle P\rangle=\frac{4(1+e^{-2kl})}{9\pi}\mu_0f\Delta_0^3J_c^2.
\label{P_StW}\end{equation} We see that in this case the leading
asymptotic AC loss term is determined by the zero order
approximation to current density distribution and does not depend
on the first order correction as in the running wave case.

\section{Discussion}
Magnetic field fluctuations are inevitable in most practical
applications of superconductors. In this work we used the Bean
critical state model to study the effect of spatially nonuniform
fluctuations of the external magnetic field. Although solutions
have been obtained only for two model situations, they help to
understand qualitatively the effect of nonuniform fluctuations in
general. The asymptotic estimates, derived for small spatially
nonuniform fluctuations using a general gauge-invariant formula
for AC losses, are the main result of our work.

For an alternating uniform magnetic field the AC losses in a
superconductor are usually expressed via the amplitude of field
variations. This is inconvenient if the external field is
nonuniform. Typically, as for the configurations considered in our
work, both the magnitude and direction of the external field
depend on position, time, problem geometry, and the fluctuation
wavelength. It seems difficult to relate the losses to any
specific characteristic of this field. The lack of a universal
direct relation between the external field at the surface of a
superconductor and AC losses becomes even more apparent if we
consider, for example, a hollow superconductive cylinder with a
long coil placed into its hole \cite{S2}. An alternating coil
current induces the shielding current in the superconductor
because the magnetic flux changes. Nevertheless, had the
superconductor been removed, the same coil current would produce
no magnetic field outside the coil at all.

Because of this reason the formulas for AC losses in this work are
presented in terms of the depth to which fluctuations penetrate
into the superconductor.  To  determine this depth and the induced
current density asymptotically for small fluctuations, we found
first the shielding surface current. Spreading this current into
the bulk and taking into account the current density constraint,
we were able to obtain the zero order approximation to current
density which was then further improved.

Such an approach is not limited to slab configuration and the two
types of fluctuations considered above. Thus, using the method
derived recently by Bhagwat and Karmakar \cite{BK1}, one can find
the surface current shielding a cylinder of an arbitrary given
cross section placed into a uniform external magnetic field.  This
makes possible to extend, following the scheme used in our work,
the asymptotic solution for cylinders in alternating uniform
transverse field \cite{Carr,Ashkin} to cylinders with non-circular
cross sections (in this case the penetration is weak if its depth
is much smaller than the characteristic cross section size).

The results obtained enable one to estimate AC losses in
superconductors of magnetic bearings and levitation systems, where
the typical configuration is similar to that of a running wave
fluctuations of external field near the surface of a thick slab
($kl\gg 1$, one-sided action of a nonuniform external field).
Suppose we can remove the superconductor and measure the
tangential component of field fluctuations at the position of slab
surface. Let us approximate this component by a running wave with
an amplitude $H_t$. By the method of images, surface current
shielding these fluctuations  has the amplitude $2H_{t}$. This can
be used to estimate the penetration depth, $\Delta_0\approx
2H_t/J_c$. The losses can now be approximated using the formula
(\ref{P_RW}) derived for the running wave fluctuations.

Numerical simulations based on a variational reformulation of the
critical-state model helped us to envision solution structures and
to control accuracy of  asymptotic solutions. The asymptotic
solutions, obtained at first by means of simple physical
arguments, were presented as zero-order terms of a consistent
asymptotic expansion. Finding the first order correction allowed
us to improve these solutions. It has been shown that the
correction ensures shielding of external magnetic field up to the
second order, $H/H_e\sim O(\nu^2)$, and (see Figs. \ref{FigAs2RW},
\ref{FigAs2SW}) provides for a satisfactory approximation for
small parameter values up to $\nu\sim 0.4$. For both the running
and standing wave fluctuations, the maximal penetration depth
$\Delta_{max}$ is smaller than the value $\Delta_0$ given by zero
order approximation: we found
$\Delta_{max}\approx\Delta_0(1-\frac{\nu}{2})$ in the first case,
$\Delta_{max}\approx\Delta_0(1-\frac{2\nu}{\pi})$ in the second
one.

Expressed via the penetration depth, AC losses for running wave
fluctuations (\ref{P_RW}) are given by exactly the same formula as
AC losses for uniform field fluctuations. We cannot expect such a
coincidence also for standing wave fluctuations, since the
penetration zone depth is not constant in that case. However, the
local losses (\ref{PlocStW}) are close if $kl\ll 1$, which simply
means that, locally, the long-wave limit is similar to the uniform
fluctuations case. If the wavelength is shorter, the losses can be
at most twice smaller than in the uniform case for the same local
penetration depth. Using the average penetration depth,
$\langle\Delta\rangle=2\Delta_0/\pi$, we can present the average
losses (\ref{P_StW}) as $\langle
P\rangle=\frac{2}{3}C\mu_0fJ_c^2\langle\Delta\rangle^3,$ where
$C=\pi^2(1+e^{-2kl})/12$. Since $0.82\le C\le 1.64$, the average
losses are also close to AC losses in the uniform field case.

For standing wave fluctuations, the leading AC loss term is
determined by zero order approximation to current density
distribution, i.e., an approximation that is comparatively easy to
find. This is similar to the case of uniform field fluctuations
considered earlier in \cite{Ashkin}. It is, therefore, surprising
that the situation is different for the running-wave type of
fluctuations: here knowledge of the first order correction to such
a solution is necessary. Neglecting this correction would cause an
error of the same order as the AC loss value itself.

To explain this discrepancy, let us note that the expressions
(\ref{P0R}) and (\ref{P0S}) for AC losses for running and standing
waves, respectively,  have different structures. In (\ref{P0S}),
the small parameter appears only in a combination
$\nu\Delta(y)/\Delta_0=\nu\Psi(y)$, where
$\Psi=\Psi_0+\nu\Psi_1+...$ We can write $P_0(y)={\cal
F}(y,\nu\Psi(y))$ and check that the function ${\cal F}(y,u)$
satisfies ${\cal F}(y,0)={\cal F}_u(y,0)=0$. Hence $P_0(y)
=\frac{\nu^2}{2}{\cal F}_{uu}(y,0)\Psi_0^2(y)+O(\nu^3)$; the main
term does not depend on $\Psi_1$.

In the running wave case, the dependance on small parameter is
different because the penetration depth and the shape of the free
boundary are separated. The equation (\ref{P0R}) can be written as
$P_0=\nu{\cal F}(\nu,\Psi),$ where ${\cal F}$ is a functional and
$\Psi=\Psi_0+\nu\Psi_1+...$ as in the previous case. Expanding we
get $P_0=\nu{\cal F}(0,\Psi_0)+\nu^2[{\cal F}_{\nu}(0,\Psi_0)+
({\cal F}_{\Psi}(0,\Psi_0),\Psi_1)]+O(\nu^3).$ Here ${\cal
F}_{\Psi}$ is the Fr\`{e}chet derivative, the first term turns out
to be zero, and so the leading term depends on $\Psi_1$.

In both cases, however, the leading term of AC losses is
proportional to $\Delta_0^3$, which corresponds to $P\sim H^3$
dependance known for the uniform field fluctuations. The next
approximation would lead to a deviation from cubic law. In the
frame of the Bean model, such deviation can be due to the shape of
a superconductor being different from that of a slab \cite{SM,S2},
heating caused by AC losses \cite{SM}, or, as in the present case,
because of spatial non-uniformity of the external field.

\appendices
\section{} The vector potential of the current
(\ref{Jmod}) and the opposite one near $x=+l$ can be written as
$$A_i=\mu_0G*J=\frac{\mu_0J_c}{4\pi}\int_{-\infty}^{\infty}
\int_0^{\Delta}S(k(y'-vt)-\Psi(\varsigma/\Delta))\ln\left(
\frac{(x+l-\varsigma)^2+(y-y')^2}{(x-l+\varsigma)^2+(y-y')^2}\right)d\varsigma\,dy'.$$
Changing the variables, $u=\varsigma/\Delta$,
$s=k(y'-vt)-\Psi(u)$, and using the Taylor expansion we obtain:
$$A_i=\frac{\mu_0J_c}{4k\pi}\Delta_0
(1+r_1\nu+...)\int_{-\infty}^{\infty}S(s) \int_0^1\ln\left(
\frac{[k(x+l)-\nu \frac{\Delta}{\Delta_0}u]^2
+[k(y-vt)-s-\Psi(u)]^2}{[k(x-l)+\nu\frac{\Delta}{\Delta_0} u]^2
+[k(y-vt)-s-\Psi(u)]^2}\right)du\,ds=$$
\begin{equation}=\frac{\mu_0h_0}{2k\pi}\left[A_0+
A_1\nu+O(\nu^2)\right]. \label{Taylor}\end{equation} Here
\begin{equation}
h_0=\frac{i_e}{2}e^{-k(a-l)}\label{h0}\end{equation} is introduced
as a  characteristic magnitude of external field fluctuations at
the surface of the superconductor and
\[A_0=\int_0^1D(x,y,t,u)du,\ \ \ \ A_1=\int_0^1u
F_1(x,y,t,u)du+ \int_0^1\Psi_1(u) F_2(x,y,t,u)du+r_1A_0,\] where
\[D=\int_{-\infty}^{\infty}S(s)\ln\left(
\frac{[k(x+l)]^2 +(\Gamma-s)^2}{[k(x-l)]^2
+(\Gamma-s)^2}\right)ds,\]
\[F_1=-2\int_{-\infty}^{\infty}S(s)\left(
\frac{k(x+l)}{(\Gamma-s)^2+[k(x+l)]^2}+\frac{k(x-l)}{(\Gamma-s)^2+[k(x-l)]^2}\right)ds,\]
\[ F_2=2\int_{-\infty}^{\infty}S(s)(\Gamma-s)\left(-\frac{1}{(\Gamma-s)^2+
[k(x+l)]^2}+\frac{1}{(\Gamma-s)^2+[k(x-l)]^2}\right)ds,\]
\[\Gamma=k(y-vt)-\Psi_0(u).\]
To calculate the integrals we present the periodic function $S$ as
the Fourier series,
\[S(s)=\sum_{n=1}^{\infty}b_n\sin(ns),\ \ \ \ b_n=\left\{\begin{array}{ll}\frac{4}{\pi n}&\
n=2m+1,\\0&\ n=2m\end{array}\right.\] and obtain
\[D=4\pi\sum_{n=1}^{\infty}\frac{b_n}{n}e^{-nkl}\sinh(nkx)\sin(n\Gamma),\]
\[ F_1=4\pi\sum_{n=1}^{\infty}b_ne^{-nkl}\sinh(nkx) \sin(n\Gamma),\ \
F_2=-4\pi\sum_{n=1}^{\infty}b_ne^{-nkl}\sinh(nkx)\cos(n\Gamma).\]
We further obtain
\[A_0=4\pi\sum_{n=1}^{\infty}
\frac{b_n}{n}e^{-nkl}\sinh(nkx)[\alpha_n\sin(nk(y-vt))-\beta_n\cos(nk(y-vt))],\]
where $b_n=0$ for all even $n$ and, for odd $n$,
\[\alpha_n=\int_0^1\cos(n\Psi_0(u))du
=\left\{\begin{array}{ll}\pi/4,& n=1,\\0 & n>1\end{array}\right. \
\ \ \ \ \ \beta_n=\int_0^1\sin(n\Psi_0(u))du=0.\] Hence,
\begin{equation}A_0=4\pi e^{-kl}\sinh(kx)\sin(k(y-vt)).\label{A0}\end{equation} Similarly,
\begin{equation}\begin{split}A_1=&4\pi\sum_{n=1}^{\infty}b_ne^{-nkl}\sinh(nkx)[\gamma_n\sin(nk(y-vt))-
\delta_n\cos(nk(y-vt))]\\ &+r_14\pi
e^{-kl}\sinh(kx)\sin(k(y-vt)),\end{split}\label{A1}\end{equation}
where all even series terms are zero and, for n odd,
\begin{eqnarray}
\gamma_n=\int_0^1u\cos(n\Psi_0(u))du-\int_0^1\Psi_1(u)\sin(n\Psi_0(u))du,\label{g}\\
\delta_n=\int_0^1u\sin(n\Psi_0(u))du+\int_0^1\Psi_1(u)\cos(n\Psi_0(u))du.\label{d}
\end{eqnarray}
We can now substitute the expressions (\ref{A0}) and (\ref{A1})
into  (\ref{Taylor}), use (\ref{AeV}) with $I_e=0$, and calculate
the total magnetic field inside the superconductor but outside the
penetration zone. We see that
$A=A_e+A_i=\frac{\mu_0h_0}{2k\pi}(\nu A_1+O(\nu^2))$. Therefore,
up to the second order terms,
\[\begin{split} H_x=& 2\nu
h_0
\{ r_1e^{-kl}\sinh(kx)\cos(k(y-vt))+\\ \ &
\sum_{n=1}^{\infty}nb_ne^{-nkl}\sinh(nkx)[\gamma_n\cos(nk(y-vt))+\delta_n\sin(nk(y-vt))]
\},\end{split}\]
\[\begin{split} H_y= & -2\nu h_0
\{ r_1e^{-kl}\cosh(kx)\sin(k(y-vt))+\\ \ &
\sum_{n=1}^{\infty}nb_ne^{-nkl}\cosh(nkx)[\gamma_n\sin(nk(y-vt))-\delta_n\cos(nk(y-vt))]
\} .\end{split}\] This proves that the current density (\ref{J0})
shields the external field up to the first order in $\nu$ and is a
zero order approximation. To nullify the first order terms of
magnetic field we will now try to satisfy the conditions
\begin{equation}\gamma_n=\left\{\begin{array}{ll}-r_1/b_1\ &\ n=1,\\0 &\
n>1,\end{array}\right. \ \ \ \mbox{and} \ \ \
\delta_n=0,\label{gd_cond}\end{equation} for  odd values of $n$.
Since $\Psi_0(u)=\arcsin(1-2u)$, we denote $\tau=1-2u$ and express
$\sin(n\Psi_0(u))$ and $\cos(n\Psi_0(u))$ in (\ref{g}), (\ref{d})
via the Chebyshev polynomials of the first and second kind,
$$T_n(\tau)=\cos(n\arccos \tau ),\ \ \
U_n(\tau)=\sin((n+1)\arccos \tau)/\sin(\arccos \tau),$$
correspondingly, \ for $\tau\in [-1,1]$. It can be shown that, for
n odd,
\[\sin(n\arcsin \tau)=(-1)^{\frac{n-1}{2}}T_n(\tau),\ \ \
\cos(n\arcsin
\tau)=(-1)^{\frac{n-1}{2}}U_{n-1}(\tau)\sqrt{1-\tau^2}.\] Defining
$\psi_1(\tau)$ as $\Psi_1(\frac{1-\tau}{2})=\Psi_1(u)$ and
calculating the integrals of known functions, we obtain
\[\begin{array}{c}\gamma_n=\frac{1}{2}(-1)^{\frac{n-1}{2}}
\left(\int_{-1}^{1}\frac{1-\tau}{2}
U_{n-1}(\tau)\sqrt{1-\tau^2}d\tau-\int_{-1}^{1}\psi_1(\tau)T_n(\tau)d\tau\right)=\\
\frac{1}{2}(-1)^{\frac{n-1}{2}}\left(\left\{\begin{array}{ll}\pi/4&\
n=1,\\0&\ n>1
\end{array}\right\}-\int_{-1}^{1}\psi_1(\tau)T_n(\tau)d\tau\right)
\end{array}\label{gamma}\]
\[\begin{array}{c}\delta_n=\frac{1}{2}(-1)^{\frac{n-1}{2}}\left(\int_{-1}^{1}
\frac{1-\tau}{2}T_n(\tau)d\tau+\int_{-1}^{1}\psi_1(\tau)U_{n-1}(\tau)\sqrt{1-\tau^2}d\tau\right)=\\
\frac{1}{2}(-1)^{\frac{n-1}{2}}\left(d_n+
\int_{-1}^{1}\psi_1(\tau)U_{n-1}(\tau)\sqrt{1-\tau^2}d\tau\right),\label{delta}\end{array}
\]
where $n$ is odd and
\[d_n=\frac{1}{2}\left[\frac{1}{(n+1)^2-1}+\frac{1}{(n-1)^2-1}\right].\] We can now rewrite
the conditions (\ref{gd_cond}) as
\begin{equation}
\int_{-1}^{1}\psi_1(\tau)T_n(\tau)d\tau=\left\{\begin{array}{ll}\frac{\pi}{4}\left(\frac{1}{2}+r_1\right)\
\ \  &n=1,\\0&n>1\end{array}\right.\label{Tcond}\end{equation}
\begin{equation}\int_{-1}^{1}\psi_1(\tau)U_{n-1}(\tau)\sqrt{1-\tau^2}d\tau=-d_n\label{Ucond}\end{equation}
and use them to determine $\psi_1(\tau)$ and $r_1$. Let us present
$\psi_1$ as the sum of its even and odd parts, $\psi_1^e$ and
$\psi_1^o$. Since functions $T_n$ are odd  for odd n,  even for
even n, and orthogonal on $[-1,1]$ with the weight
$1/\sqrt{1-\tau^2}$, condition (\ref{Tcond}) means that
$\psi_1^o(\tau)\sqrt{1-\tau^2}=cT_1(\tau)$ where $T_1(\tau)=\tau$
and
\[c={\frac{\pi}{4}\left(\frac{1}{2}+r_1\right)}
\left/\int_{-1}^1\frac{T_1^2(\tau)d\tau}{\sqrt{1-\tau^2}}\right.=\frac{1}{2}\left(\frac{1}{2}+r_1\right).\]
Thus $\psi_1^o=c\tau/\sqrt{1-\tau^2}$ and is singular at $\tau=\pm
1$ if $c\neq 0$. Since the expansion (\ref{Taylor}) is valid only
for $|\nu\Psi_1(u)|\ll 1$, we need a nonsingular function and must
set
$$r_1=-\frac{1}{2}$$ to make $c=0$. This determines the first order
correction to the penetration depth:
$$\Delta=\Delta_0(1-\frac{1}{2}\nu+O(\nu^2)).$$
The function $\psi_1$ becomes even  and we can expand it into a
series of the Chebyshev polynomials of the second kind containing
only the polynomials of even orders:
\begin{equation}\psi_1(\tau)=\sum_{m=0}^{\infty}a_mU_{2m}(\tau).
\label{amU}\end{equation} The polynomials $U_n$ are orthogonal on
$[-1,1]$ with the weight $\sqrt{1-\tau^2}$, so the coefficients of
this series are easily found from the condition (\ref{Ucond}):
$$
a_m=-d_{2m+1}\left/\int_{-1}^1U_{2m}^2(\tau)\sqrt{1-\tau^2}d\tau\right.
=-\frac{1}{\pi}\left(\frac{1}{4(m+1)^2-1}+\frac{1}{4m^2-1}\right).$$
We have now satisfied the conditions (\ref{gd_cond}) but there
appears a contradiction: although the series (\ref{amU}) converges
for $\tau\in (-1,1)$, $\psi_1$ becomes infinite at $\tau=\pm 1$.
Indeed, at these points $U_{2m}=2m+1$ and $a_{m}\approx -1/2\pi
m^2$ for big $m$. Since the conditions (\ref{gd_cond}) were
derived under assumption $|\nu\Psi_1(u)|\ll 1$, the  function
determined by this series cannot be accepted as the first order
correction to $\Psi_0$.

The computations, however, were not in vain and this singularity
can be eliminated, if we take only a finite number of terms and
define
\begin{equation}\psi_1(\tau)=\sum_{m=0}^{M}a_mU_{2m}(\tau).\label{psi1_mod}
\end{equation} It can be shown that in this case
$|\psi_1(\tau)|<\ln(M+1)/\pi$ for all $M>1$ and
$|\psi_1(\tau)|<16/15\pi$ for $M=0$ or $1$. Thus, the assumption
remains valid provided that
\begin{equation}\left\{
\begin{array}{ll}\nu\ln(M+1)\ll
\pi& \ \mbox{if}\ M>1,\\ \nu\ll\pi &\ \mbox{if}\ M=0,1.
\end{array}\right.\label{logM}\end{equation} It is easy to see
that conditions (\ref{gd_cond}) for $\gamma_n$ are still satisfied
for all $n$, whereas the conditions for $\delta_n$ hold only for
$n\le 2M+1$. For $n>2M+1$ we have
\[\delta_n=\frac{1}{2}(-1)^{\frac{n-1}{2}}d_n\approx
\frac{1}{2n^2}(-1)^{\frac{n-1}{2}}.\] The non-compensated magnetic
field inside the superconductor has, up to the second order in
$\nu$, the components
\[\begin{split}H_x&=2\nu h_0
\sum_{n=2M+3}^{\infty}nb_n\delta_ne^{-nkl}\sinh(nkx)\sin(nk(y-vt)),\\
H_y&=2\nu h_0
\sum_{n=2M+3}^{\infty}nb_n\delta_ne^{-nkl}\cosh(nkx)\cos(nk(y-vt)).
\end{split}\]
For $|x|<l-\Delta_0$ we have $e^{-nkl}\cosh(nkx)<e^{-n\nu}$,
$e^{-nkl}|\sinh(nkx)|<e^{-n\nu}$; also $nb_n=4/\pi$ for odd $n$.
Hence,
\[ |H_x|,|H_y|\lesssim\nu h_0\frac{4}{\pi}\sum_{m=M+1}^{\infty}\frac{e^{-(2m+1)\nu}}{(2m+1)^2}<
\nu
h_0\frac{4}{\pi}e^{-(2m+3)\nu}\int_{M}^{\infty}\frac{dm}{(2m+1)^2}<\nu
h_0\frac{2e^{-(2M+3)\nu}}{\pi(2M+1)}\] which proves that the field
can be made $O(\nu^2)$ if \begin{equation}
\frac{e^{-(2M+3)\nu}}{2M+1}\lesssim
O(\nu).\label{Onu}\end{equation} This means that the first order
correction to $\Psi_0(u)$ may be chosen as
$\Psi_1(u)=\psi_1(1-2u)$, where $\psi_1(\tau)$ is given by
(\ref{psi1_mod}) and $M$ satisfies the conditions (\ref{logM}) and
(\ref{Onu}).
\section{}
The vector potential of the current (\ref{JSt}) can be written as
$$
A_i=\frac{\mu_0J_c}{4\pi}\int_{-\infty}^{\infty}\int_{0}^{\Delta(y')}S(ky')
\ln\left(
\frac{(x+l-\varsigma)^2+(y-y')^2}{(x-l+\varsigma)^2+(y-y')^2}
\right)d\varsigma\,dy'.$$ Changing the variables,
$u=\varsigma/\Delta(y')$ and $s=ky'$, we obtain
\[A_i=\frac{\mu_0J_c}{4k\pi}\int_{-\infty}^{\infty}S(s)\Delta(s/k)\left\{
\int_{0}^{1} \ln\left( \frac{[k(x+l)-\nu
\frac{\Delta(s/k)}{\Delta_0}u]^2+(s-ky)^2}{[k(x-l)+\nu
\frac{\Delta(s/k)}{\Delta_0}u]^2 +(s-ky)^2} \right)du\right\}ds\]
with $S(s)\Delta(s/k)=\Delta_0(\sin(s)+\nu\Psi_1(s)S(s)+...)$.
Integrating in $u$ and expanding we get
\[A_i=\frac{\mu_0h_0}{2k\pi}[A_0+A_1\nu+O(\nu^2)],\]
where the characteristic amplitude of fluctuations $h_0$ is
defined by (\ref{h0}),
\[A_0=\int_{-\infty}^{\infty}\ln\left(\frac{[k(x+l)]^2
+(s-ky)^2}{[k(x-l)]^2+(s-ky)^2} \right)\sin (s)\, ds=4\pi
e^{-kl}\sin(ky)\sinh(kx),\]
\begin{equation}\begin{split}
A_1=&\int_{-\infty}^{\infty}\Psi_1(s)S(s)
\ln\left(\frac{[k(x+l)]^2+(s-ky)^2}{[k(x-l)]^2+(s-ky)^2} \right)\,
ds -\\ &
\int_{-\infty}^{\infty}\left\{\frac{k(x+l)}{[k(x+l)]^2+(s-ky)^2}
+\frac{k(x-l)}{[k(x-l)]^2+(s-ky)^2} \right\}\sin (s)|\sin(s)|\,
ds.\end{split}\label{A1St}\end{equation} Obviously, $\Psi_1$
should be a $\pi$-periodic and even (because of symmetry)
function. To calculate the integrals in (\ref{A1St}) we present
$\Psi_1(s)S(s)$ and $|\sin(s)|\sin(s)$ as the Fourier series,
\[\Psi_1(s)S(s)=\sum_{n=1}^{\infty}a_n\sin(ns),\ \ \ \ \
|\sin(s)|\sin(s)=\sum_{n=1}^{\infty}b_n\sin(ns),\] where
$b_n=-\frac{8}{\pi n(n^2-4)}$ for odd $n$,  $b_n=0$ for even $n$,
and the coefficients $a_n$ are unknown. Integrating, we find \[
A_1=2\pi\sum_{n=1}^{\infty}\left(\frac{2}{n}a_n+b_n\right)e^{-knl}\sin(nky)\sinh(nkx).\]
Up to $O(\nu^2)$, the total magnetic potential can be written as
\[ A=A_e+A_i=\nu\frac{\mu_0h_0}{2k\pi}A_1=
\nu\frac{\mu_0h_0}{k}\sum_{n=1}^{\infty}
\left(\frac{2}{n}a_n+b_n\right)e^{-knl}\sin(nky)\sinh(nkx).
\] To make zero the $O(\nu)$ terms of
magnetic field inside the superconductor, it is sufficient to
satisfy the conditions
\[a_n=-n\frac{b_n}{2}=\left\{\begin{array}{lc}\frac{4}{\pi(n^2-4)}\ &\
n=2m+1,\\0&n=2m.\end{array}\right.\] This gives
\[\Psi_1(s)S(s)=\sum_{m=0}^{\infty}a_{2m+1}\sin(\{2m+1\}s)\]
and so, up to the second order terms, the penetration depth can be
presented as
\[\Delta(y)=\Delta_0\{|\sin(ky)|+\nu
V(ky)\mbox{sign}[\sin(ky)]\},\] where the series
\[V(s)=\frac{4}{\pi}\sum_{m=0}^{\infty}\frac{\sin(\{2m+1\}s)}{(2m+1)^2-4}\]
can be summed up analytically.
 Let us denote $z=e^{is}$,
\[U(s)=\frac{4}{\pi}
\sum_{m=0}^{\infty}\frac{\cos(\{2m+1\}s)}{(2m+1)^2-4},\] and
consider the complex function
\[\begin{split}{\cal
W}=&U(s)+iV(s)=\frac{4}{\pi}\sum_{m=0}^{\infty}\frac{z^{2m+1}}{(2m+1)^2-4}=
\frac{1}{\pi}\sum_{m=0}^{\infty}z^{2m+1}\left[\frac{1}{2m-1}-\frac{1}{2m+3}
\right]=\\
\ &
\frac{1}{\pi}\left[\frac{1}{z}-z+\left(z^2-\frac{1}{z^2}\right)
\sum_{m=1}^{\infty}\frac{z^{2m-1}}{2m-1}\right]=
\frac{1}{\pi}\left[\frac{1}{z}-z+\left(z^2-\frac{1}{z^2}\right)
\sum_{m=1}^{\infty}\int_{0}^{z}z^{2m-2}dz\right]=\\
\ &
\frac{1}{\pi}\left[\frac{1}{z}-z+\left(z^2-\frac{1}{z^2}\right)
\int_{0}^{z}\frac{dz}{1-z^2}\right]=
\frac{1}{\pi}\left[\frac{1}{z}-z+\left(z^2-\frac{1}{z^2}\right)
\frac{1}{2}\ln\left(\frac{1+z}{1-z}\right)\right].
\end{split}\]
Since $z-1/z=2i\sin(s),\ z^2-1/z^2=2i\sin(2s)$, and
$Re\left[\ln\left(\frac{1+z}{1-z}\right)\right]=
-\ln(|\tan(s/2)|)$, 
the
imaginary part of ${\cal W}$ is
\[V(s)=-\frac{2}{\pi}\sin(s)-
\frac{1}{\pi}\sin(2s)\ln(|\tan(s/2)|).\]

\section*{Acknowledgment}
L.P. appreciates helpful discussions with Yu. Shtemler and
hospitality of the Isaac Newton Institute, Cambridge, UK, where
part of this work has been carried out in the framework of 2003
Programme ``Computational Challenges in Partial Differential
Equations".

\bibliographystyle{IEEEtran} 
\bibliography{IEEEabrv,my}
\end{document}